\documentclass[USenglish,onecolumn]{article}

\usepackage[utf8]{inputenc}
\usepackage{esvect,bm,bbm,mathtools,amsthm,amssymb,amsfonts}
\usepackage{caption}
\usepackage{subcaption}
\usepackage{multirow}
\usepackage{color} 
\usepackage[colorlinks = true,
            linkcolor = blue,
            urlcolor  = blue,
            citecolor = blue,
            anchorcolor = blue]{hyperref}
\usepackage{cleveref}
\usepackage[a4paper, margin=1in]{geometry}

\newcommand{\e}{{\rm e}}
\newcommand{\im}{{\rm i}}
\newcommand{\E}{{\mathbb E}}
\newcommand{\Pa}{{\mathbb P}}

\newcommand{\R}{{\mathbb R}}
\newcommand{\N}{{\mathbb N}}

\newcommand{\Dcal}{{\mathcal D}}

\newcommand{\Ical}{{\mathcal I}}

\newtheorem{proposition}{Proposition}[section]
\newtheorem{lemma}[proposition]{Lemma}
\newtheorem{theorem}[proposition]{Theorem}
\newtheorem{definition}[proposition]{Definition}
\newtheorem{corollary}[proposition]{Corollary}

\newtheorem{example}[proposition]{Example}

\newcommand{\Ind}[1]{\mathbbm{1}_{ \left\{ #1 \right\} }}
\newcommand{\cExp}[2]{\E\left[ #1 \mid #2 \right]}
\newcommand{\Exp}[1]{\E\left[ #1 \right]}
\newcommand{\cPro}[2]{\Pa\left[ #1 \mid #2 \right]}
\newcommand{\Pro}[1]{\Pa\left[ #1 \right]}
\newcommand{\Tra}[3][]{\mathcal{T}^{#2,#3}_{#1}}

\DeclarePairedDelimiter{\ceil}{\lceil}{\rceil}
\DeclarePairedDelimiter{\floor}{\lfloor}{\rfloor}

\newcommand{\greekvec}[1]{\mbox{\boldmath$#1$}}

\newcommand{\vtheta}{\greekvec{\theta}}

\usepackage{ulem,xcolor}

\newcommand\gsout{\bgroup\markoverwith
{\textcolor{green}{\rule[1pt]{2pt}{1.5pt}}}\ULon}
  
\begin{document}
 
\title{Dependent Defaults and Losses \\ with Factor Copula Models
\footnote{The published version is available \href{http://dx.doi.org/10.1515/demo-2017-0022}{here} and this version contains an extended appendix.
The authors would like to thank for useful comments and discussions Val\'erie Chavez-Demoulin, Pierre Collin-Dufresne, Damir Filipovi\'c, Monique Jeanblanc, and Benjamin Junge, as well as participants from the 2015 CEQURA conference on Advances in Financial and Insurance Risk Management in Munich, the 2016 conference on Dependence Modeling in Finance, Insurance and Environmental Science in Munich, and the 2017 Actuarial and Financial Mathematics conference in Brussels.
The research leading to these results has received funding from the European Research Council under the European Union's Seventh Framework Programme (FP/2007-2013) / ERC Grant Agreement n.~307465-POLYTE.
}}

\author{
Damien Ackerer \footnote{Swissquote Bank. 
E-mail: \href{mailto:damien.ackerer@swissquote.ch}{damien.ackerer@swissquote.ch}} 
\and 
Thibault Vatter \footnote{Department of Statistics, Columbia University, E-mail: \href{mailto:tv2233@columbia.edu}{tv2233@columbia.edu}}
}

\date{December 22, 2017}

\maketitle

\begin{abstract}  
{We present a class of flexible and tractable static factor models for the term structure of joint default probabilities, the factor copula models.
These high-dimensional models remain parsimonious with pair-copula constructions, and nest many standard models as special cases.
The loss distribution of a portfolio of contingent claims can be exactly and efficiently computed when  individual losses are discretely supported on a finite grid.
Numerical examples study the key features affecting the loss distribution and multi-name credit derivatives prices.
An empirical exercise illustrates the flexibility of our approach by fitting credit index tranche prices.}
\end{abstract}



\bigskip
\noindent \textbf{AMS Classification (2010):} 60E05, 60E10, 62H05, 62H20, 65T50, 91G20, 91G40, 91G60

\section{Introduction}

The modeling of dependent events over time, and of random losses, is a common challenging task in insurance, quantitative risk management, financial engineering, and reliability engineering. 
In this work we suggest a parsimonious and tractable framework to model potentially inhomogeneous and dependent default times in high dimensions.
Built upon bivariate and factor copulas, our framework allows us to compute the loss distribution for portfolios of contingent claims whose losses are discretely supported on a finite grid.
We numerically illustrate how the risk profiles of such portfolios vary when changing the copula specification, and empirically verify the flexibility of our framework by fitting credit derivative prices.

Factor copulas have been widely used to directly model dependence structure in many different fields. 
While classical static credit risk models use latent variables conditional on which the default times are independent, such approaches recover the copula ``indirectly'' by assuming simple functional relationships (e.g., often linear). 
In this paper, we use copulas to directly link such variables to default probabilities.
This more direct approach generalizes virtually all standard copula models in the credit risk literature and offers two important advantages.
First, the dependence between default times and the latent factor can be heterogeneous, that is different for each entity.
Second, numerous tractable and flexible models can be constructed with parsimony using mixtures and cascades of pair copulas.

The two standard approaches to price credit derivatives with static factor models are the exact but slow recursive method, and the fast but approximate Fourier inversion method.
We combine the best of both approaches and show how the loss distribution of complex credit portfolios can be exactly and efficiently computed.
We assume that the realized individual losses take values on a finite grid and that, conditional on the latent factor, they are independent from each others and from the default times.
We discuss factor dependent Beta-binomial distributions as a flexible way to model individual loss amounts.
The aggregate loss then takes values on a finite grid whose size increases linearly with the number of firms. The discrete Fourier transform can in turn be applied to exactly recover the portfolio loss distribution.
This allows us, for example, to compute the exact payoff distribution of credit portfolio derivatives such as tranches, collateralized debt obligations (CDO) squared, and credit index swaptions. 

We numerically explore the performance and the flexibility of our framework.
We first show that the discrete Fourier method is significantly faster than the recursive method of \cite{Andersen2003,Hull2004} especially as the number of latent factors or the loss support size increase.
We then study the impact of various dependence assumptions on the loss distribution.
Several examples on tranches and CDOs squared illustrate that the loss distribution of repackaged structured products can have dramatically different risk profiles.
We also show that the distribution choice of individual firm losses, fixing the average recovery value at default, can critically affect the portfolio loss distribution as well.

The practice of inconsistently pricing complex structured credit derivatives has been blamed for partially causing the subprime financial crisis.
We study this challenging exercise within our framework.
More precisely, we calibrate several models to market tranche prices from the North America investment grade credit index series 21.
Some of the models include stochastic loss amounts which have been fitted on the historical realized recovery rates.
We suggest a mixture of two copulas as a flexible specification mimicking two regimes. 
In a static analysis, we find that the mixture outperforms the other models, as it is the only one reproducing the prices of both the junior and senior tranches. 
Calibrating this model for all days in our sample, we further find that the parameters are stable over time. 
Furthermore, one of the correlations being almost always equal to one, we repeat the exercise by fixing it to 0.999. 
Interestingly, we find similar results, therefore achieving an almost perfect calibration to all tranches using only two parameters (i.e., the other correlation and the weight). 

\vspace{0.5cm}
In summary, the contributions of this paper are to
\begin{itemize}
\setlength{\itemsep}{0em}
\item construct high-dimensional and inhomogeneous credit risk models in a parsimonious and tractable fashion using factor copulas,
\item show that any static credit risk model equipped with the conditional independence property can be equivalently rewritten as a factor copula model,
\item model random and factor dependent individual losses on a finite grid,
\item compute efficiently the exact loss distribution of complex portfolios, such as portfolio of repackaged tranches, potentially conditional on the realization of some defaults, 
\item numerically investigate how portfolio loss distributions are affected by some dependence and loss assumptions, and
\item empirically illustrate that CDX tranches can be consistently priced with a simple model.
\end{itemize}
The proposed framework is well suited to the consistent pricing and risk-management of insurance claims and financial credit derivatives.
It could for example be used to efficiently stress test large portfolios under various scenarios, such as realized firm defaults and/or changes in the dependence structure of default times.

\vspace{0.5cm}
We now review some of the related literature.
Our approach builds on recent advances on the high-dimensional modeling of random variables. When dealing with multivariate data, copulas are attractive, allowing to model separately the marginal distributions and the dependence structure. Unfortunately, few copulas remain practically useful in high-dimensional settings, because common parametric families are often either too flexible, or not enough. An example of the former is the elliptical family, whose members have a number of parameters that grows quadratically with the dimension. Conversely, members of the Archimedean family have a small and fixed number of parameters, independently of the dimension. Recently, high-dimensional copulas using a factor structure have been constructed independently by \cite{dongpatton2013,dongpatton2015} and \cite{krupskiijoe2013,krupskiijoe2015}. Such approaches alleviate the curse of dimensionality by considering a smaller set of latent variables, conditional upon which the random variables of interest are assumed independent. Arguably the main difference between the methods presented in \cite{dongpatton2013,dongpatton2015} and  \cite{krupskiijoe2013,krupskiijoe2015} is that copulas proposed in the former can only be simulated, whereas those in the latter admit closed form expressions. In fact, it can be shown the factor copulas from \cite{krupskiijoe2013,krupskiijoe2015} are a special case of pair-copula constructions (PCCs). One of the hot topics of multivariate analysis over the last couple of years, PCCs are flexible representations of the dependence structure underlying a multivariate distribution. Introduced by \cite{bedfordcooke2001,bedfordcooke2002} and popularized by \cite{aasczadofrigessibakken2009}, PCCs are decompositions of a joint distribution by considering pairs of conditional random variables. For a given joint distribution, such a construction is not unique, but all possible decompositions can be organized as graphical structures, the so-called vines. 
Assuming the copula linking default times as in \cite{krupskiijoe2013,krupskiijoe2015}, an interesting aspect of our approach is that it nests the standard models described for instance in \cite{li2000default,vasicek2002distribution,Burtschell2005,hofert2011cdo} as special cases. 

Fourier transform techniques have been considered in \cite{Gregory2003,laurent2005basket} to approximate the loss distribution, and exact recursive algorithms have been derived in \cite{Andersen2003,Hull2004} with finite and discrete support.
To the best of our knowledge, our work is the first to combine discrete Fourier transform and losses with finite and discrete support, allowing us to study structured product such as CDO squared without simulations, unlike \cite{guillaume2009pricing,hull2010risk}.
The Fourier techniques have also been applied to price stock options and insurance claims for many years, see~\cite{embrechts1993some,carr1999option,dufresne2009fourier}.
Alternatively, in some special cases such as large homogeneous portfolios, explicit expressions have been derived to approximate the portfolio loss distribution, see~\cite{vasicek2002distribution, schloegl2005note}.

The accurate modeling of dependent defaults is particularly important for credit derivatives pricing, we refer to \cite{collin2009short} for a description of standard structured credit products.
In particular, the calibration of tranches on credit portfolios is a daunting task, which is often solved in an ad-hoc way, that is by considering a specific model for each tranche.
Significant effort have been made to develop consistent models, see \cite{giesecke2008} for a comparison between top down and bottom up approaches and \cite{Jakob2014} for a comparison of different copulas in the CreditRisk+ framework. Standard copula models had however limited empirical success and other frameworks have been developed, see \cite{hull2006valuing, albrecher2007generic, brigo2007calibration, Kalemanova2007, Cousin2008, herbertsson2008pricing, fouque2009multiname, Burtschell2009, filipovic2011dynamic, mai2014multivariate}.
In this paper, we develop bottom-up models that are both simple to calibrate and successful at reproducing all the tranche spreads.
Furthermore, while the valuation of CDO squared has been considered with simulations in \cite{hull2010risk, guillaume2009pricing}, this work is the first to derive explicitly the loss distribution of a CDO squared in a factor copula framework. 
We refer to~\cite{brigo2010credit} for a technical analysis of valuation methods for structured credit products.
Note that the same defaults dependence as in our static framework may be obtained for some homogeneous portfolio in a dynamic framework where the default times are driven by a stochastic process, see the survey by~\cite{mai2013ciid} on dynamic models.

The realized loss at default on corporate loans and bonds is known to be stochastic, volatile, and negatively correlated with the business cycle.
The recovery rates volatility and correlation with default risk is studied, for example, in \cite{altman2004default}.
The valuation of credit derivatives with random losses has also been investigated in \cite{Andersen2004,krekel2008pricing,amraoui2008optimal}.

\vspace{0.5cm}
The remainder of the paper is structured as follows. 
\Cref{sec:fc} presents the factor copula framework. 
\Cref{sec:loss} describes the construction of the individual loss amounts and the exact computation of the loss distributions. 
\Cref{sec:num,sec:emp} contain respectively the numerical analysis and an application to financial market data. 
\Cref{sec:proofs,sec:sfm} in the appendix contain the proofs and the factor copula representation of some standard models.

\section{The factor copula framework} \label{sec:fc}

We first recall how dependent default times can be constructed using a copula when the marginal default probabilities are deterministic.
We then combine factor copulas and bivariate copulas to construct high-dimensional models in which the dependence between the default times and the latent factors can easily be made inhomogeneous.
We finally show that standard factor models in the credit risk literature can always be rewritten, sometimes in simpler terms, as factor copula models.

\subsection{Default Times Construction}

We consider $N$ entities. For each $j=1,\dots,N$, let $p_{j,t}$ be a non-decreasing deterministic function for all ${0 < t < \infty}$ with $p_{j,0}=0$ and $\lim_{t \rightarrow \infty} p_{j,t}=1$. 
We define the default time $\tau_j$ of entity $j$ as follows
\[
\tau_j := \inf \{ t\ge 0 : U_j  \le  p_{j,t} \},
\]
where $U_j$ is a uniform random variable on the unit interval.
This is a standard construction of default times, see~\cite{mcneilfreyembrechts2005,bielecki2013credit}.
The function $p_{j,t}$ is equivalent to the marginal default probability of entity $j$ before any default is observed
\begin{equation}
\label{eq:condproba}
\Pro{ \tau_j \le t } = \Pro{ U_j  \le p_{j,t} } = p_{j,t}.
\end{equation}
When $p_{j,t}$ is absolutely continuous with respect to time then it is given by $p_{j,t} = 1- e^{-\int_0^t \lambda_{j,s}ds}$ for some non-negative default intensity function $\lambda_{j,s}$.
Note that, in this setup, the random vector $U=(U_1,\dots,\,U_N)$ is the only stochastic object whose probability distribution is given by a copula $C_U$.
In other words, if for any vector $(u_1,\dots,\,u_N)\in [0,1]^N$ the random vector $U \in [0,1]^N$ is such that $\Pro{U_j\le u_j} = u_j$ for each $j$, then its joint distribution is called a copula and we write
\begin{equation}
C_U(u_1,\dots,\,u_N) = \Pro{U_1\le u_1 ,\dots,\,U_N\le u_N}.
\end{equation}
The following well-known lemma shows that for $(t_1,\dots,\,t_N)\in\R_+^N$ there exists a simple expression linking joint to marginal default probabilities using the copula $C_U$ of $U$.

\begin{lemma}\label{lem:jointDefPro}
The joint default probability is given by
\begin{equation}\label{eq:jointDefPro}
\Pro{ \tau_1 \le t_1 ,\dots, \, \tau_N \le t_N } = C_U\left(p_{1,t_1} ,\dots, \, p_{N,t_N} \right).
\end{equation}
\end{lemma}

A direct construction of high-dimensional copulas amounts at trading-off model complexity and tractability. This is somewhat problematic, because the usual parametric families contain either too many (e.g., in the case of implicit copulas extracted from known multivariate distributions), or too few (e.g., in the case of Archimedean copulas built using a continuous and nonincreasing $N$-monotone generator) parameters. Furthermore, as we will show in \Cref{sec:loss} when pricing complex financial derivatives, the notion of conditional independence (on a set of latent factors) allows us to obtain a flexible yet tractable class of models. Hereinafter we therefore focus on the so-called factor copulas of \cite{krupskiijoe2013,krupskiijoe2015}.

\subsection{One-factor copulas}\label{subsec:one-factor}

A one-factor copula model is constructed by assuming that there exists a latent factor $V$ uniformly distributed on the unit interval such that, conditional on the realization of $V$, the coordinates of the random vector $U$ are independent. In other words, we have
\begin{equation} \label{eq:VcondJDP}
\Pro{U_1\le u_1 , \dots, \, U_N \le u_N \mid V=v} = \prod_{j=1}^N \cPro{U_j\le u_j}{V=v}
\end{equation}
for any vector $(u_1,\dots,\,u_N)\in [0,1]^N$ and $v\in[0,1]$. 
Note that we directly consider a uniformly distributed factor. This is without loss of generality using the probability integral transform for factors having alternative continuous distributions.
While factor models in the credit risk literature usually build upon real-valued random variables, this creates an unnecessary layer of complexity layer when computing expressions such as the joint default probability~\eqref{eq:jointDefPro}.

The following proposition shows that our assumption about $V$ yields a simple decomposition in terms of bivariate copulas for $C_U$, also known as one-factor copula.
\begin{proposition}[One-factor copula \cite{krupskiijoe2013,krupskiijoe2015}]\label{prop:1fcm}
For $j=1,\dots,\,N$, let $C_{U_j,V}$ denote the joint distribution of $U_j$ and $V$, that is $\Pro{U_j\le u_j, \, V \le v} = C_{U_j,V}(u_j, v)$. If the coordinates of $U$ are independent conditionally on $V$, then 
\begin{equation}\label{eq:1fcm}
C_U(u_1,\dots,\,u_N)  = \int_{[0,1]} \;  \prod_{j=1}^N \; C_{U_j \mid V}\left( u_j\mid v\right) \, dv,
\end{equation}
where, for all $j=1,\dots,\,N$, $C_{U_j \mid V}\left( u_j \mid v \right) = \frac{\partial C_{U_j,V}(u_j,v) }{\partial v}$ are the so-called $h$-functions.
\end{proposition}
The $h$-functions have been introduced by \cite{aasczadofrigessibakken2009} while studying the pair-copula decomposition of a 
general multivariate distribution: if $C_{U_j,V}(u_j,v)=\Pro{U_j\le u_j,V\le v}$, then $C_{U_j \mid V}\left( u_j \mid v \right) = \cPro{U_j\le u_j}{V= v}$. 

Note that $C_{U_j,V}(u,v)=uv$ implies $C_U(u_1,\ldots,u_N) = \prod_{j=1}^N u_j $. In other words, if $U_j$ is independent from $V$, then it is also independent from $U_k$ for all $k\neq j$, which means that the coordinates of $U$ depend on each other only through the factor $V$. 

\begin{example}\label{ex:Li}
The Gaussian model of \cite{li2000default} is a one-factor copula model obtained by using for all $j$ the bivariate copula
$
{C_{U_j,V}(u_j,v;\rho)}={\Phi_2\left(\Phi^{-1}(u_j),\Phi^{-1}(v);\rho \right)}
$ which implies
$
{C_{U_j \mid V}(u_j \mid v;\rho)} = {\Phi\left(\frac{\Phi^{-1}(u_j)-\rho \Phi^{-1}(v)}{1-\rho^2} \right)}
$ and
$
{C_{U}(u_1, \dots, u_N;\rho)}={\int^1_{0} \prod_{j=1}^N  \Phi\left(\frac{\Phi^{-1}(u_j)-\rho \Phi^{-1}(v)}{1-\rho^2} \right) \, dv}
$
where $\Phi(\cdot)$ is the standard normal distribution and $\Phi_2(\cdot,\cdot;\rho)$ is the bivariate normal distribution with correlation $\rho$. 
\end{example}
The specification in \Cref{prop:1fcm} is more flexible than standard factor models from the credit risk literature. The reason is that it is straightforward to build non-homogeneous models with firm-specific dependence between the default time and the common factor. For instance, one could build a model using a different bivariate copula for each conditional probability $\Pa[U_j\le u_j \mid V=v]$. Furthermore, there exists many well-studied parametric families for bivariate copulas, see~\cite{Schepsmeier2014}.

Beyond such parametric families, a simple way to increase the modeling flexibility while preserving analytical tractability is to combine different bivariate copulas. 
Hence, we suggest mixture distributions as a natural and simple extension, and which considerably enrich the class of one-factor copulas models.

\begin{definition}
Let $K$ be a positive integer, $C_{U_j, V}$ is a mixed bivariate copula if there exists $K$ copulas $C_{U_j, V}^k$, $K$ positive weights $w_k>0$ such that $\sum_{k=1}^K w_k=1$, and
\begin{equation}\label{eq:copmix}
C_{U_j, V}(u_j,v) = \sum_{k=1}^K w_k C_{U_j, V}^k(u_j,v).
\end{equation}
\end{definition}
One way to interpret this expression is Bayesian, namely assuming that the dependence between the random variable $U_j$ and the factor $V$ is uncertain and follows the distribution $C_{U_j, V}^k$ with probability $w_k$.
The corresponding $h$-function still has a simple expression, as we have ${C_{U_j \mid V}(u_j \mid v)} = {\sum_{k=1}^K w_k C_{U_j \mid V}^k(u_j \mid v)}$. 
While a mixture of Gaussian copulas was studied in~\cite{li2005cdo}, the definition above can accommodate a different parametric family for each mixture component.

The loss distribution of a credit portfolio conditional on the default time of a specific entity is of particular interest for risk management applications. For instance, it is a necessary input to compute a Credit Valuation Adjustment (CVA), namely the expected loss on a bilateral position resulting from the default risk of this entity (see \cite{zhu2007guide} for an introduction). As it turns out, the joint distribution of default times conditional on a subset of realized default times is obtained as a simple modification of Equation~\eqref{eq:1fcm}.

Let $\Ical=\{1,\dots,\,N\}$ and $\Dcal \subset \Ical$ denote respectively the entire set and a subset of entities.
The following proposition shows that the joint default distribution conditional on the defaults of all the entities in $\Dcal$ also has a simple representation.
\begin{proposition}\label{prop:fcmdens}
In a one-factor copula model, the joint default distribution conditional on $\tau_k=t_k$ for $k\in \Dcal$ is
\begin{multline} \label{eq:1fcm:condP}
\Pro{ \tau_1 \le t_1 ,\dots, \, \tau_N \le t_N \mid \tau_k=t_k : k\in \Dcal} = \\ 
\frac{\int_{[0,1]} \; \prod_{j\in \Ical \setminus \Dcal} C_{U_j \mid V}\left( p_{j,t_j}\mid v\right) \; \prod_{k\in \Dcal}  c_{U_k , V}\left( p_{k,t_k}, v\right)  \, dv}{\int_{[0,1]} \; \prod_{k\in \Dcal}  c_{U_k , V}\left( p_{k,t_k}, v\right)  \, dv}
\end{multline}
where $c_{U_j , V} (u,v) = \frac{\partial^2 C_{U_j , V}(u,v)}{\partial u \partial v}$
is the density of the bivariate copula $C_{U_j , V}$.
\end{proposition}
Although the default times are correlated, conditioning on a subset of defaulted entities does not significantly complexify the expression for the joint distribution of the surviving entities. The denominator on the right hand side in \eqref{eq:1fcm:condP} is the copula density of the defaulted entities evaluated at the default times.
%

\subsection{Multi-factor copulas}

We generalize the framework of \Cref{subsec:one-factor} by considering a $d$-dimensional random vector of latent factors $V=(V_1, \dots, V_d)$. 
We assume that $V$ takes values on the hypercube $[0,1]^d$ and has uniform marginal distributions.
The joint distribution of $V$ is by definition a copula that we denote $C_V$. 
The following proposition shows that the one-factor framework extends to a multi-factor one.
\begin{proposition}[Multi-factor copula]\label{prop:fcm}
For $j=1,\dots,\,N$, let $C_{U_j,V}$ denote the joint distribution of $U_j$ and $V$, that is $\Pro{U_j\le u_j, \, V \le v} = C_{U_j,V}(u_j, v)$. If the coordinates of $U$ are independent conditionally on $V$, then
\begin{equation}\label{eq:fcm}
C_U(u)  = \int_{[0,1]^d} \prod_{j=1}^N \; C_{U_j \mid V}\left( u_j\mid v\right)\,d C_V(v)
\end{equation}
where, for all $j=1,\dots,\,N$,  $
 C_{U_j \mid V}(u_j\mid v) = \frac{\partial^d C_{U_j, V}(u_j,v)}{\partial v_1\dots\,\partial v_d}$.
\end{proposition}

Note that the representation in \eqref{eq:fcm} is not equivalent to the one found in \cite{krupskiijoe2013,krupskiijoe2015} since the coordinates of $V$ are not necessarily independent. 

The representation~\eqref{eq:fcm} is arguably more complicated than \eqref{eq:1fcm} despite their apparent similarity. 
The reason is that, instead of being bivariate, each $C_{U_j,V}$ has dimension $d+1$. 
However, the multi-factor framework simplifies under the assumption of independent latent factors $V$ as shown in the following proposition.
We denote the function composition with the symbol $\circ$, that is $f \circ g(x) = f(g(x))$ for any real valued functions $f$ and $g$. 

\begin{corollary}[Copulas with independent factors \cite{krupskiijoe2013,krupskiijoe2015}]\label{cor:fcmPCC} 
If $C_V(v) = \prod^d_{j=1} v_j$, then
\begin{equation}\label{eq:PCC}
C_{U}(u_1,\,\dots,\,u_n) = \int_{[0,1]^d} \prod_{j=1}^N C_{U_j\mid V_1}(\cdot|v_1) \circ \cdots \circ C_{U_j\mid V_d}(u_j|v_d) dv,
\end{equation}
where $C_{U_j, V_k}$ is a bivariate copula for $j \in 1,\dots,\,d$ and $k=1,\dots,\,d$.
\end{corollary}

Note that the recursive decomposition \eqref{eq:PCC} is a particular case of pair-copula constructions \cite{krupskiijoe2013,krupskiijoe2015}.
This construction is interesting for several reasons. First, it is a parsimonious way to model a complex multivariate dependencies. Second, the hierarchical structure, which can be represented as a graphical model, has an intuitive interpretation. Third, because the integrand in \eqref{eq:PCC} is a simple recursion, it can be vectorized in a computationally efficient manner.

Finally, it should be noted that the number of latent factors is also the dimension of the hypercube on which the product of conditional copulas has to be integrated to compute the joint default probability. One should therefore balance modeling flexibility and computational cost.

\subsection{Comparison with standard factor models}

We show that standard static factor models can be rewritten explicitly as factor copula models. In this context, one usually considers a random vector $ Y = (Y_1,\,\dots,\,Y_N)\in\R^N$ along with a deterministic and componentwise non-decreasing vector  $ y_{t}= (y_{1,t},\,\dots,\,y_{N,t})\in\R^N$. For instance, $Y$ and $y_t$ (or their exponentials) can represent the firm values and corresponding default barriers.
The default time $\tau_j$ of firm $j$ is then defined as the first time its value is below its default barrier, that is $\tau_j = \inf\{ t\ge 0 : Y_{j}\le y_{jt}\}$.
Additionally, standard factor models are constructed by decomposing the stochastic behavior of the firm value into a systemic and an idiosyncratic component. In other words, one assumes the existence of a random vector $X\in\R^d$ and $N$ variables $\epsilon_j$ for $j \in \left\{ 1, \dots, N \right\}$, such that $Y_j$ is a function $X$ and $\epsilon_j$, that is $ Y_j = f_j(X,\epsilon_j)$ for some $(d+1)$-dimensional function $f_j$ taking values on $\R_+$.

Let $F_{Y_j},\,F_X$, respectively $F_{Y_j}^{-1},\,F_X^{-1}$, denote the distributions of $Y$ and $X$, respectively their inverse, and $F_{Y_j\mid X}$ denote the conditional distribution of $Y_j$ given $X$. 
The following proposition shows that any standard factor model is equivalent to a specific factor copula model.

\begin{theorem}\label{thm:fm2fcm}
A standard factor model is a factor copula model with marginal default probabilities $ p_{j,t}=F_{Y_i}(y_{j,t})$ and conditional copulas
\begin{equation*}
C_{U_j\mid V} (u \mid v) = F_{Y_j \mid X}( F^{-1}_{Y_j}(u) \mid (F^{-1}_{X_1}(v_1),\,\dots,\,F^{-1}_{X_N}(v_N)),
\end{equation*}
for $j=1,\,\dots,\,N$, and where the copula of $V$ is given by
\begin{equation*}
C_V(v) = F_X(F^{-1}_{X_1}(v_1),\,\dots,\,F^{-1}_{X_N}(v_N)).
\end{equation*}
Furthermore, if the functions $F_X$ and $F_{Y_j}$ for all $j=1,\dots,N$ are continuous, then the copulas $C_V$ and $C_{U_j\mid V}$ for all $j=1,\dots,N$ are unique.
\end{theorem}

While $C_{U_j\mid V}$ and $C_V$ sometimes admit closed-form expressions, it is clear that the marginal distributions are irrelevant. Instead, working directly with copulas offers more modeling flexibility while ensuring tractability.

\begin{example}\label{ex:Li2}
The Gaussian model described in \Cref{ex:Li} is obtained by writing, for $j \in \left\{ 1, \dots, N \right\}$, $Y_j = \rho X + \sqrt{1-\rho^2} Z_j$
and $y_{j,t} = \Phi(p_{j,t})$
where $X,Z_1, \dots, Z_N$ are i.i.d. $N(0,1)$ random variables.
\end{example}

In \Cref{sec:sfm} we derive the factor copula representation of other popular models. 


\section{Discrete loss distributions} \label{sec:loss}

We first define a specific finite grid on which losses will take values.
Then, we show that portfolio loss distributions can be computed in quasi-closed form using discrete Fourier inversion, and that this methodology also applies to the underlying portfolio of complex credit derivatives such as CDO squared.
The section concludes by presenting a flexible approach to model random and factor dependent loss amounts. 

\subsection{Loss Amounts Specification}

We define the time-$t$ loss $L_t$ on a portfolio composed of securities written on $N$ different entities,
\begin{equation}\label{eq:Ltdef}
L_t = \sum_{j=1}^N \ell_j \, \Ind{\tau_j \leq t} = \sum_{j=1}^N \ell_j \, \Ind{U_j \leq p_{jt}},
\end{equation}
where $\ell_j$ is the possibly random loss amount experienced when entity $j$ defaults, and $\Ind{\tau_j \leq t}$ is the default indicator of entity $j$. In this section, we make two assumptions on $\ell_j$ to preserve the tractability of the portfolio loss distribution, and to enable efficient numerical techniques.

First, we assume that $\ell_j$ is $V$-conditionally independent of both $\ell_k$ for $k \neq j$ and $U$ (or equivalently $\tau$), that is 
\begin{equation*}
\Pro{ U \le u , \, \ell \le x \mid V = v} = \prod_{j=1}^N  C_{U_j\mid V}(u_j \mid V = v) \Pro{\ell_j \le l_j\mid V = v},
\end{equation*}
with $\ell=(\ell_1,\dots,\,\ell_N)$, and for any $u\in[0,1]^N$, $v\in[0,1]^d$ and $l\in\R^N_+$.
As for the joint distribution of default times, $V$-conditional probabilities can be arbitrarily specified. 
Hence, this assumption does not preclude dependent default times and losses given default.
Note that, in the literature, loss amounts are commonly assumed to be independent from each others and from the default times, and often set to be constant.
Our framework goes beyond such limitations, which can be of practical importance as shown in~\Cref{sec:num}

Second, we assume that losses have a discrete distribution with finite support. More specifically, we let $\delta\in\R_+$ be the common loss unit, such that each $\ell_j$ has a discrete and finite support starting at zero and with mesh $\delta$, that is
\begin{equation*}\label{eq:lgdss}
\ell_j \in \{0, \, \delta,\, 2\delta,\,\dots,\, m_j  \delta  \}, \quad j=1,\,\dots,\,N
\end{equation*}
for some integer $m_j\in\N$. Hence, the portfolio loss distribution also has a discrete support with the same mesh $\delta$, that is
\begin{equation*}
L_t\in \left\{0, \, \delta,\, 2\delta,\,\dots,\, M  \delta \right\}
\end{equation*} 
where $M=\sum_{i=1}^N m_i$. Although $\delta$ is an arbitrary constant, it can be as fine as required in order to mimick the discreteness of real-world prices. For instance, assuming that the granularity of prices is in cents (i.e., $\delta=0.01\$$) and that the notional of each contract is 1\$, then $m_j=100$ and $M=N\times100$.
This setup can also be understood as a specific discretization of continuously distributed random losses. The use of a common loss unit can be traced back at least to~\cite{Andersen2003,Hull2004}. 

\subsection{Portfolio loss distribution}

We show that the portfolio loss distribution has an almost closed-form expression that can be efficiently computed numerically. Recall that, for a discrete and finitely supported random variable $X \in \left\{0, 1, \dots, M \right\}$ admiting a characteristic function $\phi_X(u)=\Exp{\e^{\im u X}}$, its distribution can be represented as a finite sum 
\begin{align*}
\Pro{X = k} = \frac{1}{M+1}\sum^M_{m=0} \phi_X\left(\frac{2\pi m}{M+1} \right)e^{- \frac{2\pi \im km}{M+1}},
\end{align*}
where $\im$ denotes the imaginary unit.
Therefore, if the characteristic function of the loss distribution admits a closed-form expression, so does the loss distribution itself. Using the $V$-conditional independence, the following proposition shows that the characteristic function of the loss admits a simple expression. To improve the clarity of the formulas, we work with the normalized losses ${\ell_j\delta^{-1} \in \left\{0, 1, \dots,  m_j \right\}}$
and normalized portfolio loss ${L_t\delta^{-1} \in \left\{0, 1, \dots,  M \right\}}$.
\begin{proposition}\label{prop:CICFformula}
The characteristic function of the normalized portfolio loss $L_t\delta^{-1}$ is given by
\begin{equation}\label{eq:cfci}
\phi_{L_t}(u) = \Exp{\e^{\im u L_t  \delta^{-1}}} = \int_{[0,1]^d} \; \prod^N_{j=1} \left(1 - p_{j,t}(v) + p_{j,t}(v) \phi_{\ell_j}(u,v) \right) \, d C_V(v),
\end{equation} 
for any time $t\ge 0$ and $\text{for } u\in\R$, where $p_{j,t}(v) = C_{U_j\mid V}(p_{j,t} \mid v)$ is the conditional default probability of $j$, $p_{j,t}$ is the unconditional default probability of $j$ in~\eqref{eq:condproba}, and 
\begin{equation}\label{eq:CFellj}
\phi_{\ell_j}(u,v) = \sum_{k=0}^{n_j} \cPro{\ell_j = \delta \, k}{V=v} \e^{\im u k }
\end{equation}
the $V$-conditional characteristic function of $\ell_j\delta^{-1}$.
\end{proposition}

The characteristic function is therefore explicit, up to the integral over the compact set $[0,1]^d$ which can be efficiently computed using, for example, Legendre quadrature. Denoting by $\left\{ w_i, x_i\right\}_{i=1}^n$ $n$ pairs of quadrature weights and nodes to approximate an integral over $[0,1]$, they can be combined in a straigthforward way to perform multi-dimensional integration by using a product rule. In other words, \Cref{eq:cfci} can be approximated by
\begin{align*}
\phi_{L_t}(u) \approx \; \sum_{i_1, \cdots, i_d} \prod^d_{l=1} w_{i_l}  \prod^N_{j=1} \left(1 - p_{j,t}(v_{i_1, \cdots, i_d}) + p_{j,t}(v_{i_1, \cdots, i_d}) \phi_{\ell_j}(u,v_{i_1, \cdots, i_d}) \right) \, d C_V(v_{i_1, \cdots, i_d}),
\end{align*}
where $v_{i_1, \cdots, i_d} = (x_{i_1}, \cdots, x_{i_d})$.
Even though the number of quadrature nodes and weights can be chosen small, such an approach works well only with a small number of factors, as the total number of grid points is $n^d$. If $d$ is larger than 4 or 5, sparse grids (see e.g., \cite{heiss2008}) can be used to reduce the number of grid points required.

The following Lemma is a reminder that, since the support of $L_t$ is discrete and finite, its distribution is equal to the discrete Fourier transform (DFT) of its characteristic function.

\begin{lemma}\label{lem:cf2pmf}
The probability distribution of the portfolio loss is given by
\begin{equation}\label{eq:cf2pmf}
\Pa\left[ L_t = k \, \delta\right] = \frac{1}{M+1} \sum_{m=0}^{M} \phi_{L_t}\left( \mu m\right) \e^{-\im \mu k m  } \mbox{ for } k\in\{0,\, \dots, \,M\},
\end{equation}
with $\mu=2\pi/(M+1)$ and $\phi_{L_t}(\cdot)$ is the characteristic function of $L_t\delta^{-1}$.
\end{lemma}
Calculating directly $\Pa\left[ L_t = k \, \delta\right]$ is in general a combinatorial problem whose complexity is increasing exponentially fast with $M$. 
However the complexity of computing the DFT in \Cref{lem:cf2pmf} is $\mathcal{O}(M^2)$ which is of significant practical importance as long as evaluating the characteristic function is also fast. Note that the Fast Fourier Transform (FFT) algorithm, whose complexity is only $\mathcal{O}(M\log(M))$, can also be used.

The assumption of loss unit and discretely supported portfolio losses appears already in \cite{Andersen2003,Andersen2004,Hull2004}, where the distribution is computed without approximation by a recursive algorithm. However, as will be shown in \Cref{subsec:comput}, the computational cost of this recursion increases much faster with both the support size and the number of factors than that of our approach.
Note also that the discrete Fourier inversion in \Cref{lem:cf2pmf} differs from the continuous Fourier inversion described in \cite{laurent2005basket,Burtschell2009} which aims to approximate a continuous loss distribution. Since our approach provides quasi-closed expressions for the loss distribution, its scope is much wider.

Whereas tranches and CDO squared can be priced using the portfolio loss distribution only (see \Cref{sec:tranchesCDO2}), other products necessitate the joint distribution of the total number of defaulted entities and of the total loss. This is the case for credit index swaptions, namely options on indices paying realized losses in exchange for premium payments proportional to the number of non-defaulted entities, the market for which is currently booming.
We now derive this joint distribution.
Let the number of defaulted entities at time $t$ be
\begin{equation}\label{eq:Ntdef}
N_t=\sum_{j=1}^N \Ind{\tau_j\le t}.
\end{equation}
The following Proposition gives a generic expression for the joint distribution of $(N_t,L_t)$.

\begin{proposition}[] \label{prop:NtLtpmf}
The joint distribution of $(N_t,L_t)$ is given by
\begin{equation}\label{eq:NtLtpmf}
\Pro{N_t = n , \, L_t = \delta k}=  \sum_{j=0}^N \;\sum_{l=0}^M \; \frac{ \phi_{N_t,L_t}(\mu j, \nu l) \, e^{-\im  \mu n j } e^{-\im   \nu k l } }{(1+N)(1+M)}
\end{equation}
with $\mu=2\pi/(M+1)$, $\nu=2\pi/(N+1)$, and 
\begin{align*}
\phi_{N_t,L_t}(x,y) = \int_{[0,1]^d} \; \prod^N_{j=1} \left(1 - p_{j,t}(v) + p_{j,t}(v) \phi(x,y,v) \right) \, d C_V(v)
\end{align*}
where $p_{j,t}(v)$ is as in \Cref{prop:CICFformula}, and $\phi(x,y,v) = \sum_{k=0}^{n_j} \; \cPro{\ell_j = \delta \, k}{V=v} \e^{\im (x + y k)}$.
\end{proposition}
The computation of~\eqref{eq:NtLtpmf} thus boils down a two-dimensional discrete Fourier transform inversion.
Note that $\phi(x,y,v)$ is the $V$-conditional characteristic function of $x + y \ell_j \delta^{-1}$ evaluated at one, that is ${\phi(x,y,v)=\E [\exp(\im(x + y \ell_j \delta^{-1}))\mid V=v]}$
and $\phi_{N_t,L_t}(x,y)$ is the characteristic function of $(N_t,L_t\delta^{-1})$ evaluated at $(x,y)$, that is ${ \phi_{N_t,L_t}(x,y) = \E [\exp(\im(xN_t + yL_t\delta^{-1}))]}$.

When the loss amounts $\ell_j$ are homogeneous and independent from $V$, then the following more direct calculation can be applied
$$\Pro{N_t = n , \, L_t = \delta k} = \cPro{L_t=k\delta}{N_t=n}\Pro{N_t = n } $$
where $\Pro{N_t = n }$ can be computed as in \Cref{lem:cf2pmf}, and ${\cPro{L_t=k\delta}{N_t=n} = \Pa[\sum_{j=1}^n \ell_j = k\delta]}$ 
can also be derived using the discrete Fourier transform.

Similar expressions can be derived for the default probabilities when the default intensities are driven by a stochastic process, see for example \cite{schonbucher2001copula}. However these expressions involve expectations of copula functions in random marginal default probabilities whose computations generally require costly numerical techniques.
Yet, combining the linear credit risk models presented in~\cite{ackerer2016linear} and polynomial factor copulas would result in tractable polynomial models with dependent default times and stochastic default intensities.
In that case, the joint default probability rewrites as an integral over the expectation of a polynomial in a polynomial diffusion which is an analytical expression, see~\cite{filipovic2016polynomial}.
Some examples of polynomial copulas, such as the Farlie-Gumbel-Morgenstern copula, can be found in \cite{nelsen1999introduction} and the Bernstein copulas, which can approximate any copula, are studied in \cite{sancetta2004bernstein}.

\subsection{Tranches and CDO squared}\label{sec:tranchesCDO2}

We describe how to compute loss distributions for two structured credit products: tranches and portfolios of tranches, also known as CDO squared.
To the best of our knowledge, this is the first time that a methodology to retrieve the exact loss distribution of a CDO squared is presented for an inhomogeneous bottom-up model.

A tranche on a credit portfolio is a derivative that pays a fraction of the realized portfolio losses above the attachment point $a$ and below the detachment point $b$ with $0\leq a < b $, in exchange of regular payments on the remaining size of the  tranche.
Define the tranche loss as
\begin{align}\label{eq:tranchedef}
\Tra[t]{a}{b} &:= \min\left\{\max\left\{ L_t-a,\,0\right\}, \, b-a\right\},
\end{align}
and denote $\epsilon_a:= \delta - (a \mod \delta) $.
The following expresses the relationship between the probability distribution of $L_t$ and that of $\Tra[t]{a}{b}$.

\begin{proposition} \label{prop:tranche}
The tranche loss $\Tra[t]{a}{b}$ probability mass function is given by
\begin{align*}
& \Pro{\Tra[t]{a}{b}=\epsilon_a + k\delta} = \Pro{L_t=(k+\ceil{a/\delta})\delta},  \quad \text{for any $k\in\N$ such that $ 0< \epsilon_a + k\delta < b-a$,}\\
&\Pro{\Tra[t]{a}{b}=0} = \sum_{m=0}^{\floor{a/\delta}} \Pro{L_t=m\delta}, \quad \text{and} \quad 
\Pro{\Tra[t]{a}{b}=b-a} = \sum_{m=\ceil{b/\delta}}^{M} \Pro{L_t=m\delta},
\end{align*}
where $\floor{x}$ (respectively $\ceil{x}$) denotes the closest integer smaller (respectively larger) than $x$.
\end{proposition}

The CDO squared loss distribution can also be computed explicitly, even when the default times of entities composing the different portfolios are dependent.
Let us consider $K$ tranches on portfolios written on different entities. For each portfolio $k$ we denote by $\Tra[k,t]{a_k}{b_k}$ and $L_{k,t}$ the $k$-th tranche and portfolio loss, with $a_k$ and $b_k$ the $k$-th tranche attachment and detachment points.

Denote  $\mathcal{L}_{t} = \sum_{k=1}^K \Tra[kt]{a_k}{b_k}$ the CDO-squared loss, at time $t$.
Assume that for all $k$, we have
\begin{align}\label{eq:abModEll}
a_k \mod \delta = 0 \quad \text{and} \quad  b_k \mod \delta = 0.
\end{align} 
Then, each of the tranche as well as the CDO squared have discrete and finite supports, that is $\Tra[kt]{a_k}{b_k} \in \left\{0,\,\delta,\,2\delta, \,\dots,\, b_k-a_k \right\}$ for  each $k$ and ${\mathcal{L}_{t} \in \left\{0,\,\delta,\,2\delta, \,\dots,\, M_K \delta \right\}}$
where $M_K=\sum_{k=1}^K (b_k-a_k)/\delta$.

\begin{corollary}\label{cor:CDOsq}
If~\eqref{eq:abModEll} holds for $k=1,\dots,K$ , then the characteristic function of the squared loss is
\[
\phi_{\mathcal{L}_{t}}(u) = \int_{\R^d} \prod_{k=1}^K \phi_{\mathcal{T}_{kt}}(u,v) dC_V(v),
\]
where
 \[
\phi_{\mathcal{T}_{kt}}(u,v) = 
\sum_{n=1}^{(b_k-a_k)/\delta} \cPro{\Tra[kt]{a_k}{b_k}= n\delta}{V=v} \e^{\im u n},
\]
is the $V$-conditional characteristic function of $\mathcal{T}_{kt}\delta^{-1}$.
\end{corollary}
To compute $\phi_{\mathcal{T}_{kt}}$ one may use \Cref{prop:tranche} applied to the $V$-conditional portfolio loss distribution, namely $\cPro{L_{kt}=m\delta}{V=v}$.
Applying \Cref{lem:cf2pmf} with $\mathcal{L}_{t}$ replacing $L_t$, one finally obtains the distribution of the squared loss.
With the distribution of the squared loss, one can then price derivatives such as tranches on a portfolio of tranches.

\subsection{Beta-binomial loss amounts}\label{sec:bb}

We describe how to use the Beta-binomial in order to model dependent loss amounts.
More specifically, we allow the distribution of the loss amounts for each entity to be dependent on the default times and others loss amounts.  
For each $j=1,\dots,N$, we let the loss amount $\ell_j$ take value in a set of the form
\[
\ell_j \in \{ b_j \delta,(a_j+b_j) \delta,\dots, (n_ja_j+b_j )\delta \} \subset \{0,\delta,2\delta,\dots,m_j\delta\}
\]
with the integers  $a_j,b_j,n_j\in\N$ such that $n_j a_j+b_j =m_j >0$.
Note that the two sets are equivalent when $a_j=1$ and $b_j=0$.
The Beta-binomial model is obtained by assuming that the $V$-conditional distribution of the loss amount increment $(\ell_j\delta^{-1}-b_j)/a_j$ is a Beta-binomial random variable.
We recall that the Beta function is defined by ${\rm B}(\alpha,\,\beta)=\Gamma(\alpha)\Gamma(\beta)/\Gamma(\alpha+\beta)$ with the Gamma function $\Gamma(z)=\int_0^\infty x^{z-1}\e^{-x}dx$.

\begin{definition}[The Beta-binomial loss model]
The $V$-conditional probability of loss is
\begin{align*}
\cPro{ \ell_j = (a_j k+b_j) \delta }{V=v} &= \int_{[0,1]} \cPro{Z=k}{p,n_j} \pi_j\left(p\mid V=v\right)dp
\end{align*}
for any $k=0,\dots,\,n_j$, where $\cPro{Z=k}{p,n_j}   = {n_j\choose k}p^k(1-p)^{n_j-k}$, and with the Beta distribution $\pi_j\left(p\mid V=v\right) =  p^{\alpha(v)-1} (1-p)^{\beta(v)-1}/\mathrm{B}(\alpha(v),\,\beta(v))$
for some functions $\alpha:[0,1]^d\rightarrow \R_{+*}$ and $\beta:[0,1]^d\rightarrow \R_{+*}$.
\end{definition}
Conditional on $V$ the number of loss units experienced upon default is the sum of a constant $b_j$ and of $k$ units $a_j$ where $k$ follows a Binomial distribution with parameter $p$ and support $0,\dots,\,n_j$.
In addition, the probability $p$ is random and distributed according to a Beta distribution with parameters $\alpha(v)$ and $\beta(v)$.
Note that the functions $\alpha$ and $\beta$ may be entity specific.

The Beta-binomia is a flexible distribution that nests a large spectrum of distributions such as the Bernouilli (see below), the discrete uniform (when $\alpha=\beta=1$), and asymptotically the binomial (for large $\alpha$ and $\beta$). Its probability mass function is given by
\begin{align*}
\cPro{ \ell_j = (a_j k+b_j) \delta }{V=v} = & {n_j \choose k} \frac{{\rm B}(\alpha(v)+k,\,\beta(v)+n_j-k)}{{\rm B}(\alpha(v),\,\beta(v))} .
\end{align*}
for any $k=0,\dots,n_j$ and where $\Gamma$ denotes the gamma function.
The $V$-conditional loss amount mean and variance therefore also have a explicit expression $ \cExp{\ell_j }{V=v} = \big( a_j\frac{n_j\alpha(v)}{\alpha(v)+\beta(v)} + b_j\big)\delta$ and ${\rm Var}\left[\ell_j  \mid V=v\right] = \frac{n_j \alpha(v)\beta(v)(\alpha(v)+\beta(v)+n_j)}{(\alpha(v)+\beta(v))^2(\alpha(v)+\beta(v)+1)} {a_j}^2 \delta^2$.
Remark that the mean loss amount is positively correlated with $V$ when the function $v \mapsto \alpha(v)/(\alpha(v)+\beta(v))$ is increasing on $[0,1]$.

\begin{example}[Bernouilli model]
The loss amount distribution reduces to a Bernoulli when $n_j=1$ with probability $p(v) = \frac{\Gamma(\alpha(v)+\beta(v))}{\Gamma(\alpha(v))}  \frac{\Gamma(1+\alpha(v))}{\Gamma(1+\alpha(v)+\beta(v))}$ which can take any value in $(0,1)$ and thus also be arbitrary close to the Dirac delta function.
\end{example}

\begin{example}[Linear Beta-binomial model] \label{ex:linearBB}
Assume that $d=1$ and that the functions $\alpha,\,beta$ are linear such that
$\alpha(v) = m_1 + m_2 v$ and $\beta(v) = m_3 + m_4 v$
where $m_i>0$ for all $i=1,\dots,\,4$.
This specification is discussed in further details in \Cref{ssec:stocorloss}.
\end{example}

\section{Numerical analysis}\label{sec:num}

In this section, we compare the computational costs of the discrete Fourier transform (DFT) and recursive methods, and numerically study the loss distribution of selected factor copula models with different features.

\subsection{Computational performance}\label{subsec:comput}

\begin{figure}
\begin{center}
\includegraphics{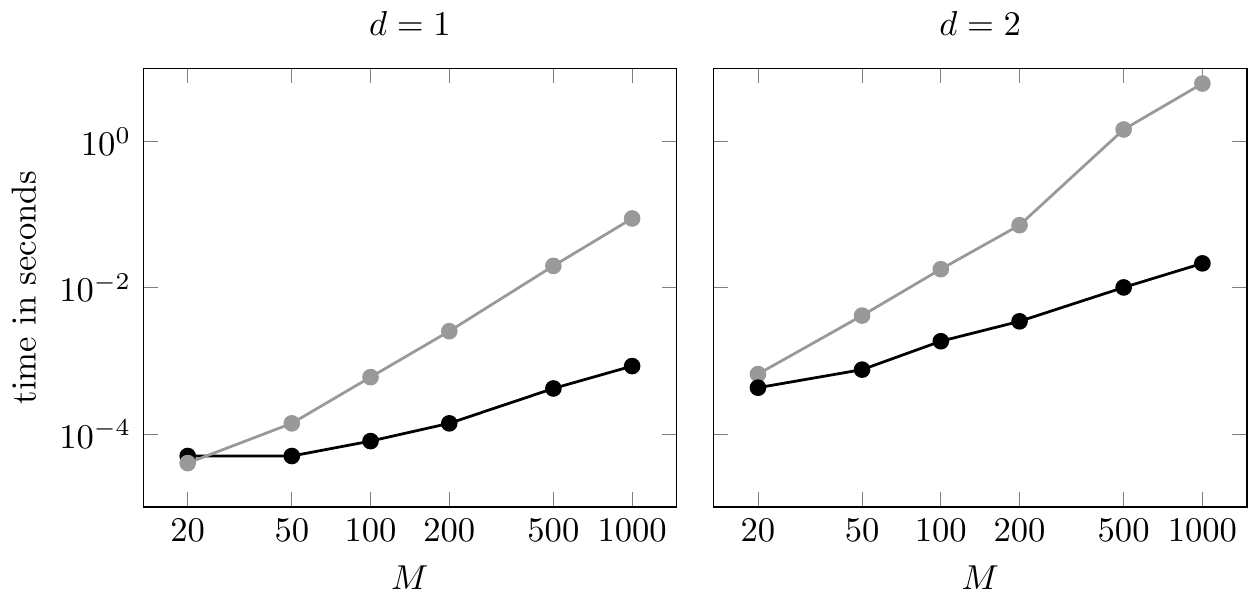}
\end{center}
\caption{Computation performance. \label{fig:DFTvsRec}}
\footnotesize
The time in seconds to compute the loss probability mass function is displayed against the loss support size $M$ for the discrete Fourier transform (black line) and recursive (grey line) methods.
The one-factor (left panel) and two-factor (right panel) standard Gaussian copula have been used under the assumption of constant loss given default. 
\end{figure}

In this section, we show that the DFT method presented in \Cref{sec:loss} is significantly more efficient than the recursive methods suggested in \cite{Andersen2003,Hull2004}.
We consider the standard one-factor and two-factor copula models, \Cref{fig:DFTvsRec} displays the computing time necessary to retrieve the probability mass function with the DFT and with the recursive method.
The calculations have been performed on a single CPU from a standard personal computer in the \textsf{R} programming language.
The DFT method is significantly faster than the recursive method in both cases: it takes roughly the same amount of time to retrieve a distribution with 1000 points with DFT and a 100 points with recursion.

\subsection{Dependent defaults with a mixed copula} \label{sec:copmix}

\begin{figure}
\begin{subfigure}[b]{\textwidth}
\begin{center}
\includegraphics{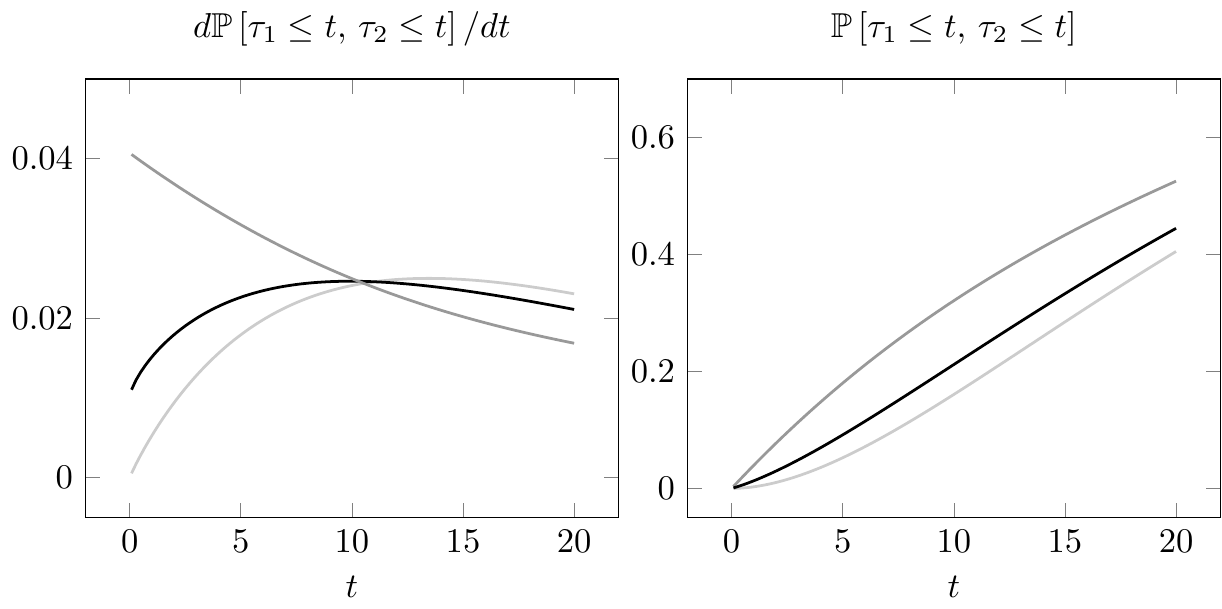}
\end{center}
\caption{Joint default probability. \label{fig:probmix}}
\end{subfigure}
\begin{subfigure}[b]{\textwidth}
\begin{center}
\includegraphics{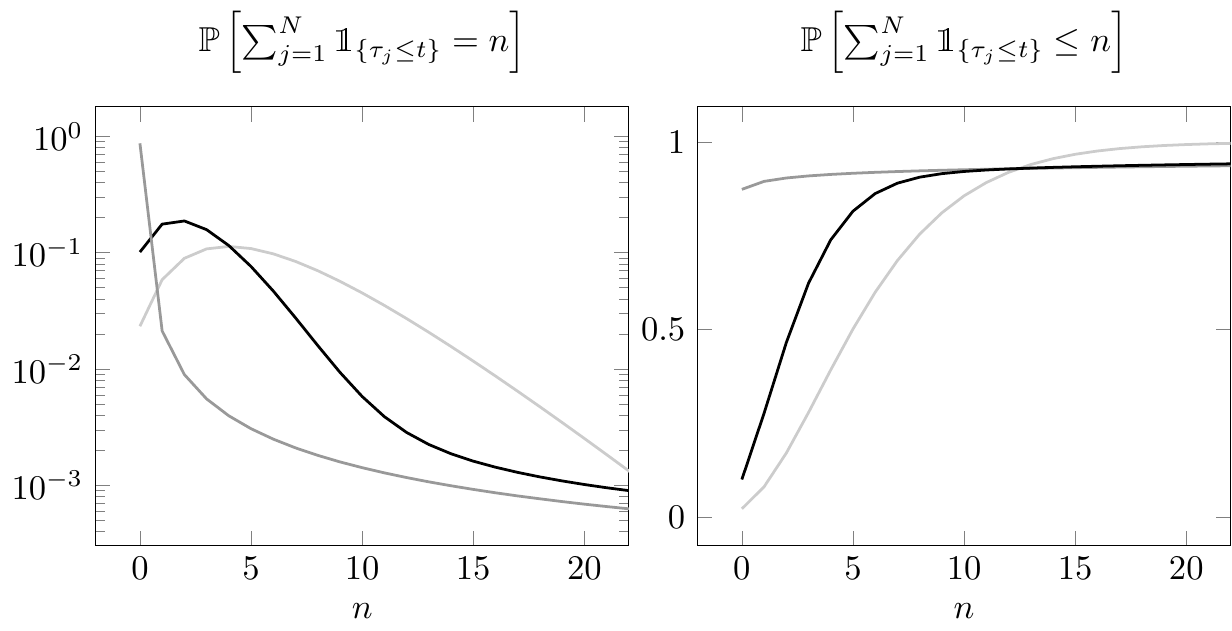}
\end{center}
\caption{Number of defaults.\label{fig:numbmix}}
\footnotesize
\end{subfigure}
\caption{Dependent defaults with a copula mixture.\label{fig:copmix}}
\footnotesize
The probability (left) and cumulative (right) density functions of the joint default of two homogeneous entities are displayed on \Cref{fig:probmix} for time horizons ranging from 1 week to 20 years for three different one-factor models.
The probability (left) and cumulative (right) density functions of the total number of defaults on a portfolio of 125 homogeneous entities are displayed on \Cref{fig:numbmix} at a 5-years horizon.
The marginal default probability is given by $p_{j,t}=0.05$ for all $j\in\Ical$ and the reference factor copula models is an equiweighted copula mixture (black line) between a Gaussian copula with $\rho=0.25$ (light-grey line) and a Clayton copula with parameter equal to 5 (grey line).
\end{figure}

We investigate the joint default probability and the total number of defaults density in a one-factor copula model with a mixed bivariate copula specification as defined in Equation~\eqref{eq:copmix}.
Fix $K=2$ and assume that $p_{j,t}=1-e^{-\lambda t}$ for $j=1,2$ with $\lambda=0.05$. 
Consider the following copula mixture
\[
C_{U_j,V}(u_j,v) = w C^{\rm C}_{U_j,V}(u_j,v) + (1-w)C^{\rm G}_{U_j,V}(u_j,v)
\]
for $j=1,\dots,N$, for some $w\in[0,1]$, and where $C^{\rm C}$ denotes the Clayton copula with parameter $5$ and $C^{\rm G}$ the Gaussian copula with parameter $0.25$.
\Cref{fig:probmix} displays the probability and cumulative density functions of joint defaults of two entities for the times $0\le t\le 20$, and for the weights $w\in\{0,0.5,1\}$.
The two limit cases therefore correspond to the Gaussian and Clayton copulas.
We observe that the joint probability of default is also a mixture of the two limit cases.
With $N=125$, \Cref{fig:numbmix} displays the total number of defaults at a 5-years horizon. 
It is clear that the distribution of the number of defaults is a mixture of the two limit components: it has the bump of the Gaussian with parameter $0.25$ and the fat tail of the Clayton with parameter $5$. 

\subsection{Credit derivatives}

We explore the loss distribution of a large portfolio, a tranche on this portfolio, and a portfolio of tranches when the underlying tranches are independent and when they depend on a common factor.
Let $N=1000$ and assume that $\ell_j=1$ and $\lambda_{jt}=0.01$ for all $j\in\Ical$ and $t\ge0$.
The reference model is the standard one-factor Gaussian copula with correlation parameter equal to $0.25$.
All the tranches have attachment point $a_k=100$ and detachment point $b_k=200$. 
The CDO squared is composed of 10 tranches to have the same loss support as the portfolio.

\begin{figure}
\begin{center}
\includegraphics{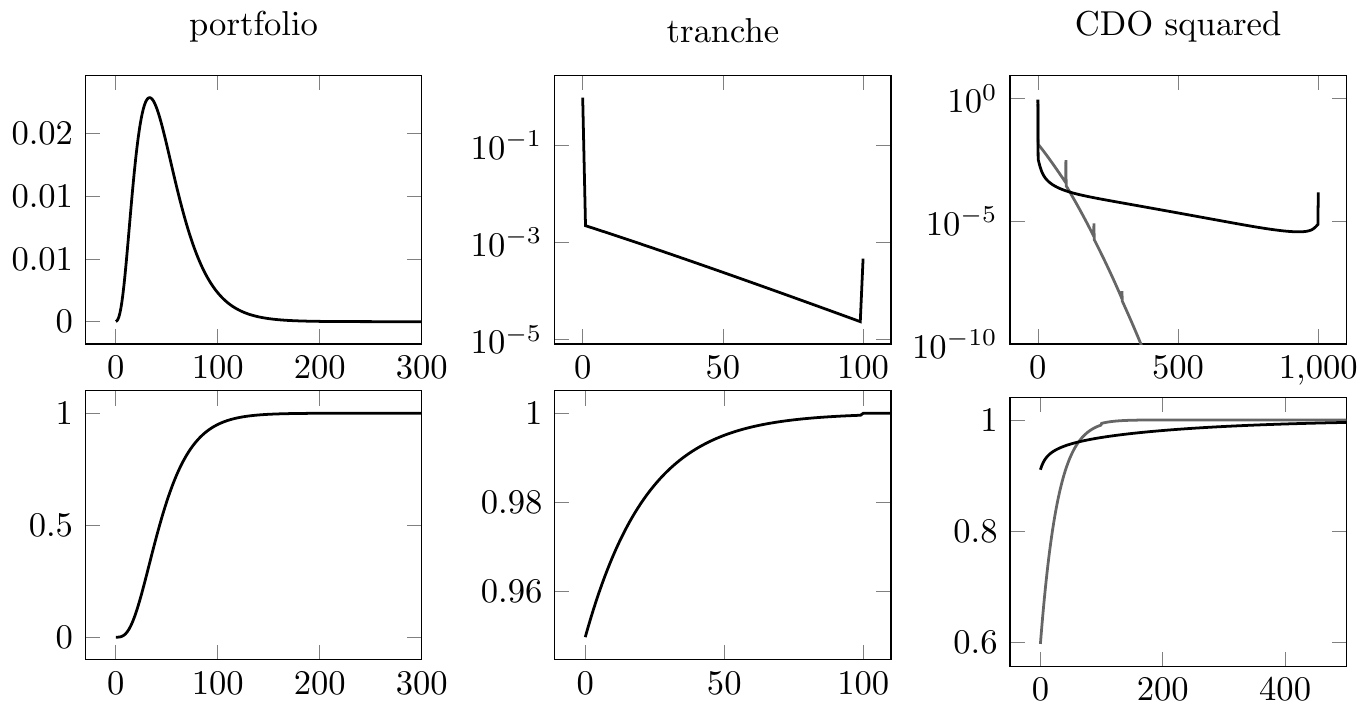}
\end{center}
\caption{Multi-name credit derivatives losses.\label{fig:loss}}
\footnotesize
The probability (first row) and cumulative (second row) density functions of the loss distribution are displayed for three different derivatives.
The first column is concerned with a portfolio of 1000 entities, the second column with a tranche on this portfolio with attachment point 100 and detachment point 200, and the third column with a portfolio of 10  such  tranches coming from different portfolios with a unique risk factor (black line) and with independent risk factors (grey line).
\end{figure}

\Cref{fig:loss} displays the probability and cumulative mass functions of the portfolio, tranche, and portfolio of tranches at the 5-year horizon.
The tranche loss distribution has two masses at the beginning and end of its support corresponding the probabilities of no loss and full loss respectively.
These concentrated masses combine to create a spiky pattern in the portfolio of tranches loss distribution.

The CDO squared loss distribution computed under the assumption of unique factor and tranche specific factor have dramatically different profiles.
With independent factors the CDO squared appears less risky than the initial portfolios.
For example, senior tranches on the pooled portfolio are virtually riskless.
On the other hand, with a unique common factor, the CDO squared has a fat tailed loss distribution and a large probability, about 91\%, of having zero losses: when the risk driver behind all tranches is the same, the diversification benefit almost completely disappears. 
\cite{hull2010risk} discussed similar results using Monte Carlo simulations.

\subsection{Stochastic and correlated loss amounts} \label{ssec:stocorloss}

\begin{figure}
\begin{center}
\includegraphics{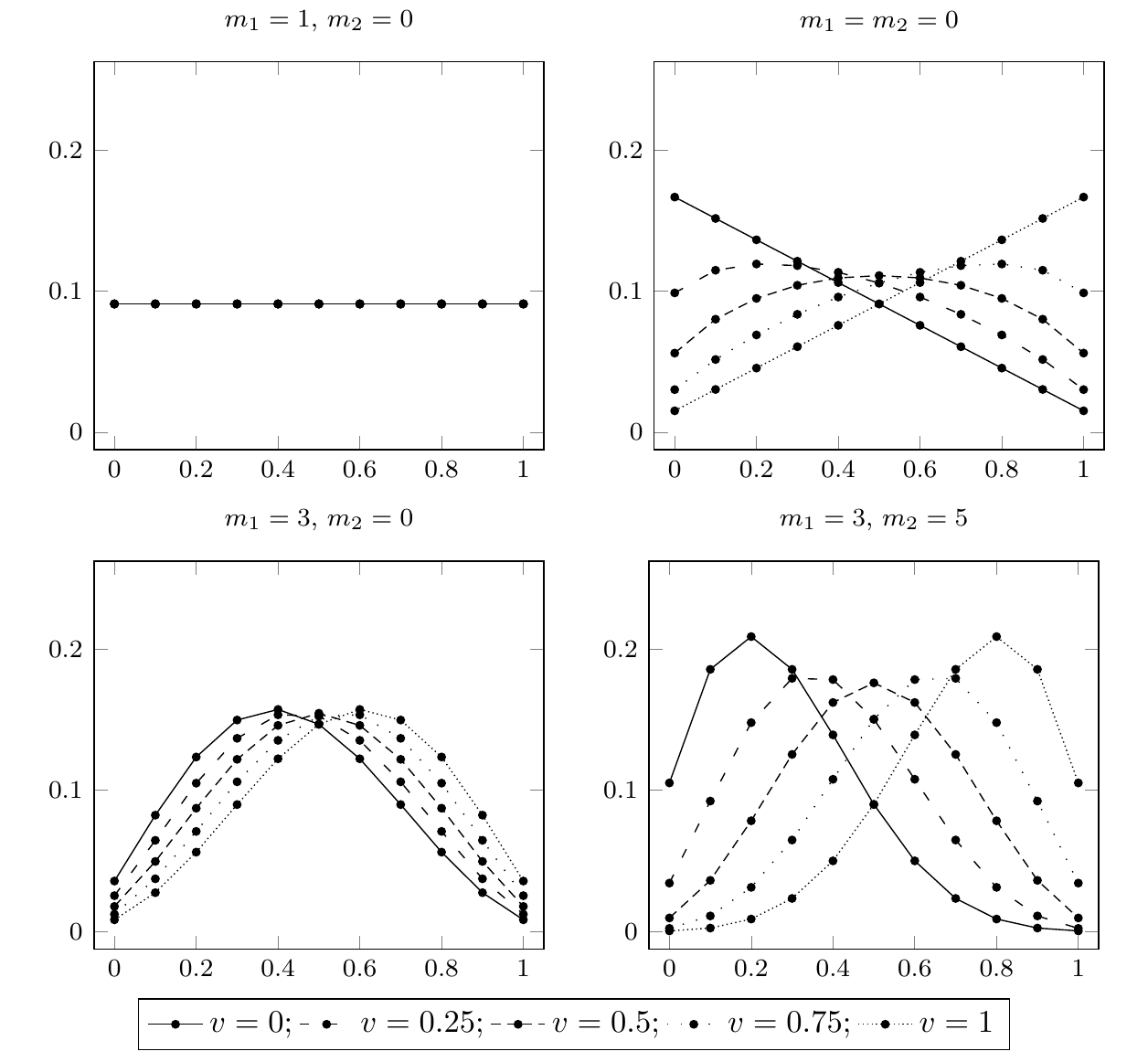}
\end{center}
\caption{Beta-Binomial distribution.\label{fig:betabinom2}}
\footnotesize
The distribution of the Beta-Binomial random variables with $n=10$ is displayed for different parameters choices.
The Beta-Binomial loss distribution for a single entity: $m_1 = 1$ and $m_2 = 0$ (top left),  $m_1 = 1$ and $m_2 = 1$ (top right), $m_1 = 3$ and $m_2 = 1$ (bottom left),  $m_1 = 3$ and $m_2 = 5$ (bottom right). In each panel, the distribution is represented for different values of the systematic factor. 
\end{figure}

In this section, we investigate the impact on the portfolio loss distribution of stochastic losses correlated with the common factor.
We consider the linear Beta-Binomial model presented in \Cref{sec:bb} with $a_j=1$, $b_j=0$, $n_j\delta=1$, $m_3=m_1$ and $m_4=m_2$.
Since
\[
\cExp{\ell_j}{V=v} =  \frac{(m_1+ m_2(1 - v))}{2m_1 + m_2},
\] 
this specification implies that the expected loss is always equal to 0.5.
\Cref{fig:betabinom2} shows that the $V$-conditional distribution of the loss amount exhibits various shapes, even-though the expected loss is constant.

\begin{figure}
\begin{center}
\includegraphics{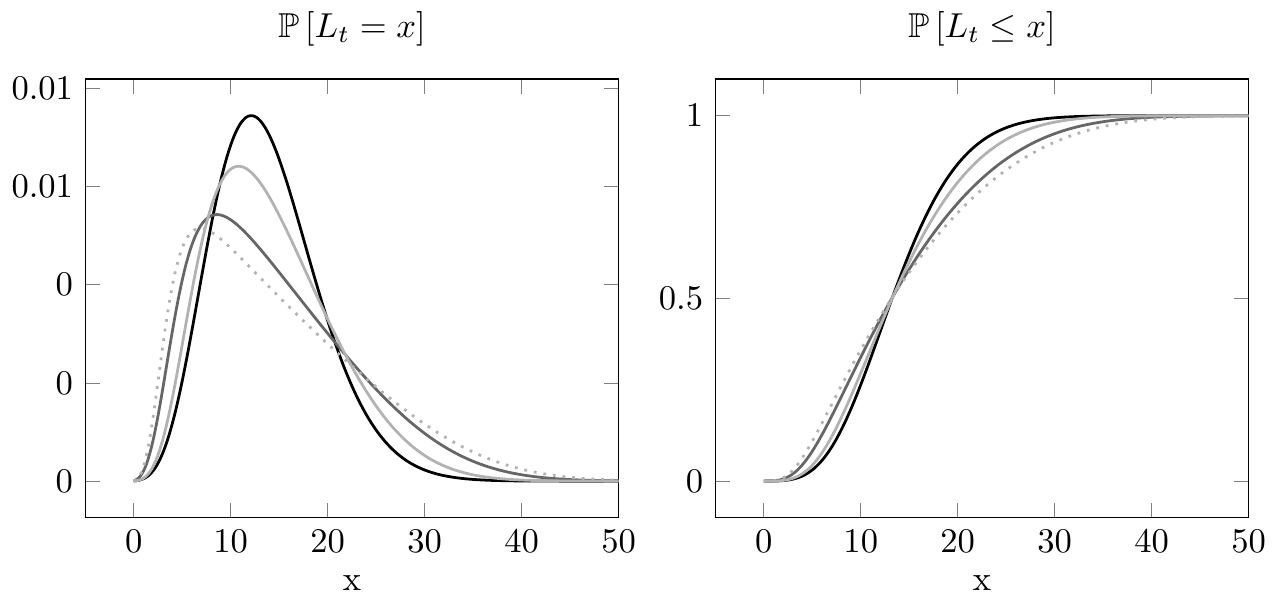}
\end{center}
\caption{Loss distribution and loss amounts dependence.\label{fig:lossBB}}
\footnotesize
The probability (left) and cumulative (right) density functions of the loss distribution are displayed for three different loss amounts specifications.
With a standard one-factor copula model with parameter equal to $0.25$, constant marginal default intensity $\lambda_{jt}=0.05$, and a 5-year horizon we consider the linear Beta-Binomial loss amounts with $m_1=m_3$ and $m_2=m_4$ for the values: 
$m_1=1$ and $m_2=0$ (black line), $m_1=1$ and $m_2=1$ (grey line), $m_1=3$ and $m_2=1$ (light-grey line), and $m_1=3$ and $m_2=5$ (dotted light-grey line).
\end{figure}

Consider the standard one-factor Gaussian copula with parameter equal to $0.25$, $N=125$, $\lambda_j=0.05$ for all $j\in\Ical$, and with the same loss amount model as above having an expected loss one half. 
\Cref{fig:lossBB} shows that the loss distribution is significantly affected by the choice of the parameters and the value of the common factor.
Compared to the benchmark case of independent and equi-distributed loss amounts (i.e., when $m_2 = 0$), increasing the dependence on the factor $V$ also increases the portfolio average loss and tail risk. 

\subsection{Number of defaults and loss dependence}

We investigate how individual losses affect the portfolio loss distribution given a number of realized defaults.
We consider the one-factor homogeneous Gaussian copula with parameter equal to $0.25$, default intensities $\lambda_{jt}=0.05$, $N=125$ entities, and for a 5-year horizon.
Let the $V$-conditional loss amounts be equal to 1 with probability $1-v$ and equal to 0 with probability so that $\E[\ell_j]=0.5$ for all $j$.

\begin{figure}
\begin{center}
\includegraphics{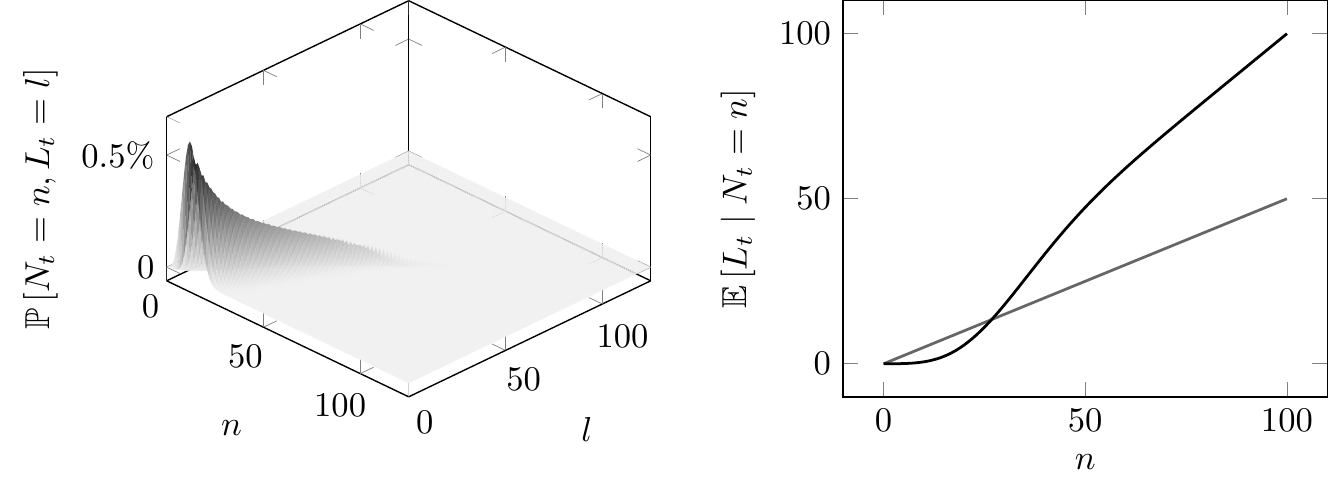}
\end{center}
\caption{Number of defaults and loss dependence.\label{fig:PMFjoint}}
\footnotesize
The left panel displays the joint probability density of the number of default $N_t$ and the loss $L_t$ when the loss amounts depend on $V$.
The right panel displays the expected loss given $n$ defaults, when the loss amounts are respectively $V$-independent (grey line) and $V$-dependent (black line).
The reference model is a one-factor homogeneous Gaussian copula with parameter equal to $0.25$, default intensities $\lambda_{jt}=0.05$, a 5-year horizon, and contains $N=125$ entities. 
The loss amount $\ell_{jt}$ is zero or one and has an expected value of 0.5.
\end{figure}

The left panel of \Cref{fig:PMFjoint} displays the density of the joint distribution of the number of defaults and loss $(N_t,L_t)$ computed as described in \Cref{lem:jointDefPro}.
We observe that most of the probability mass is concentrated on a diagonal band near the origin, and that there is little to no mass on the off diagonal parts. This is intuitive as more defaults implies larger losses.

The right panel of \Cref{fig:PMFjoint} displays the expected loss given a certain number of default, that is $\Exp{L_t\mid N_t=n}$ for $n=0,\dots,\,125$, with the black line corresponding to the model above and the grey line to $V$-independent loss amounts with $\Pro{\ell_j = 0}=\Pro{\ell_j = 1}=0.5$.
This value is increasing with $N_t=n$ as the losses are expected to increase with the total number of defaults.
Several interesting observations can be made.
First, the marginal rate of losses starts from almost zero at the origin and increases rapidly. Second, the conditional expected loss converges to the maximal possible loss.
This is contrasted with $V$-independent loss amounts where the relation between $N_t$ and $L_t$ is linear.

\section{Empirical analysis}\label{sec:emp}
In this section, we illustrate our approach by calibrating various factor copula models to credit index tranche prices. 

\subsection{Data}
We focus on tranches of the CDX.NA.IG index, which is composed of 125 investment grade North American compagnies. Historically, all tranches (except the most junior) were unfunded meaning that they were traded without upfront payments.
However, since 2009 and the resolution of the financial crisis, these products are now traded with standardized quarterly payments and the variations in financial risks are reflected in the upfront payments.
Based on liquidity and risk, new series with tenors of 3, 5, 7, and 10 years are determined in March and September.

\begin{figure}
\begin{center}
\includegraphics{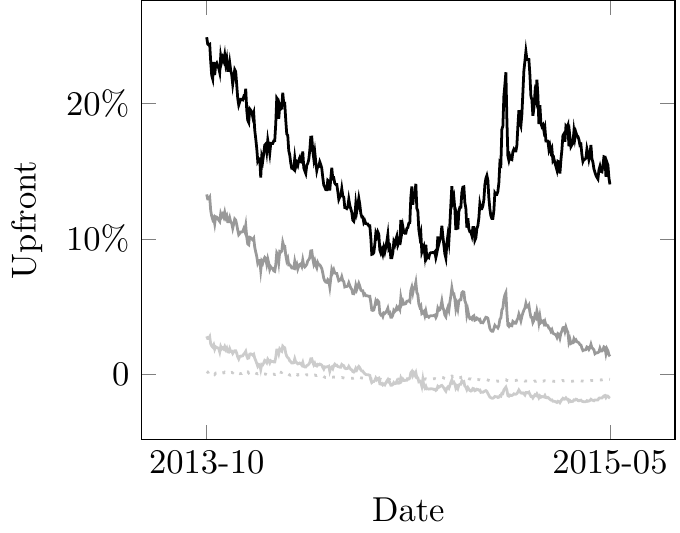}
\end{center}
\caption{CDX.NA.IG.21 upfront values in percents over the sample period.\label{fig:CDXupfronts}}
\footnotesize
\Cref{fig:CDXupfronts} displays the upfronts for the equity (black), mezzanine (grey), senior (light-grey) and super-senior (dotted light-grey) tranches.
\end{figure}

The series 21, issued in September 2013 with a tenor of 5 years, came along with four standardized tranches, whose spreads and attachments/detachments points are detailed in \Cref{tab:CDXstruct}. 
Our sample contains 405 daily upfront payments for the four tranches, which we summarize in \Cref{tab:CDXstats} and display in \Cref{fig:CDXupfronts}. 
This series was selected because it was the most recent with more than a year of existence in our sample.
By convention, the market quotes upfronts in percentage of the corresponding tranche width, which is about thirty times larger for the super-senior than for the equity.
Furthermore, the sign of the upfront is also interesting: since it is negative, one most often receives money to buy protection on the super-senior tranche, as well as on the senior tranche at the end of the sample period.

\begin{table}
\begin{subtable}[t]{\textwidth}
\begin{center}
\begin{tabular}{cccc}
 \hline
   Name &       Attachement & Detachment & Spread   \\
          \hline
Equity & 0    &    300   &     500   \\
Mezzanine & 300     &   700    &    100    \\ 
Senior & 700     &   15\%    &   100     \\
Super-senior & 15\%   &   100\%   &   25  \\
 \hline
\end{tabular}
\end{center}
\caption{Attachment-detachment points and the tranche spreads in basis points (or percents) per annum.\label{tab:CDXstruct}}
\end{subtable}

\bigskip

\begin{subtable}[t]{\textwidth}
\begin{center}
\begin{tabular}{ccccc}
  \hline
 & Equity & Mezzanine & Senior & Super-senior \\ 
  \hline
Mean & 15.18\% & 592 & -29 & -23 \\ 
  Vol & 410 & 274 & 127 & 22 \\ 
  Min & 859 & 134 & -207 & -53 \\ 
  Max & 24.87\% & 13.28\% & 282 & 25 \\ 
   \hline
\end{tabular}
\end{center}
\caption{Statistic on the upfront payments in basis points (or percents) of the tranche width.\label{tab:CDXstats}}
\end{subtable}
\caption{Structure and summary statistics on CDX.NA.IG.21 tranches.}
\end{table}

\begin{figure}
\begin{center}
\includegraphics{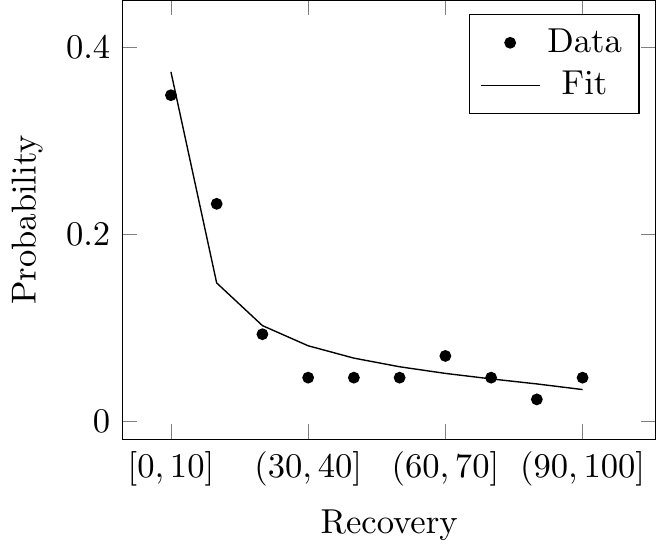}
\end{center}
\caption{Recovery data and Beta-binomial fit. \label{fig:PMFbb_realdata}}
\footnotesize
The data points represent the number of credit events leading to losses in each 10\% interval divided by the total number of events, and the solid line is the result of a Beta-binomial fit.
\end{figure}

To calibrate recoveries, we looked at all credit events leading to losses (i.e., bankruptcy, failure to pay, and restructuring) between 2005 and 2014 from entities listed in Series 1 to 22 in the CDX.NA.IG, CDX.NA.HY, iTraxx.Eur, and iTraxx.Eur.Xover. \Cref{tab:recov} presents summary statistics of the 43 credit events leading to losses. One noteworthy observation is that both the mean and median are significantly below the 40\% recovery rate commonly assumed when modeling corporate credit portfolios. Furthermore, while about half of the losses are related to CDX.NA.HY entities, which are notoriously more risky, this observation is robust to the removal of such events. In \Cref{fig:PMFbb_realdata}, we show the discretized distribution of the recoveries as well as the result of a Beta-binomial fit. The $[0\%,100\%]$ interval was divided into 10 equal-sized bins and each of the 43 recovery rates was assigned to a bin. Then, Beta-binomial parameters of $\alpha=0.4$ and $\beta=1.1$ were obtained by maximizing the likelihood, leading to a fitted mean and standard deviation of $29\%$ and $27\%$.
\begin{table}[ht]
\centering
\begin{tabular}{rrrrrrrr}
  \hline
 & Min. & 1st Qu. & Median & Mean & 3rd Qu. & Max. & Std \\ 
  \hline
\% & 0 & 6 & 16 & 28 & 42 & 94 & 28 \\ 
   \hline
\end{tabular}
\caption{Summary statistics for the recoveries \label{tab:recov}}
\end{table}

\subsection{Calibration}
Let $P^{a_i,b_i}$, $a_i$ and $b_i$ for $i \in \left\{1, \dots, 4\right\}$ denote the quoted upfronts, and attachments/detachments points of each tranches. For such derivatives, the contract buyer pays predefined coupons to the seller at the payments dates $0= T_0 \le \dots\le T_n=T$ where $T$ is the contract maturity, we call this series of cash-flow the premium leg $V_{\rm prem}$.
The contract seller pays default contingent cash-flows to the buyer at the defaults dates when losses materialize, we call this series of cash-flow the protection leg $V_{\rm prot}$.
The contract value for the buyer is then given by $V_{\rm prot}-V_{\rm prem}$.

For a model parametrized with $\vtheta \subseteq \Theta \subseteq \mathbb{R}^l$ (i.e., $l$ is the number of parameters), we denote by $P^{a_i,b_i}(\vtheta)$ the model price, that is the quantity satisfying
\begin{align*}
P^{a_i,b_i}(\vtheta)(b_i - a_i) + V^{a_i,b_i}_\text{prem}(\vtheta)  = V^{a_i,b_i}_\text{prot}(\vtheta),
\end{align*}
where $b_i - a_i$ is the tranche width, and the premium and protection legs are defined as
\begin{align*}
V^{a_i,b_i}_\text{prem}(\vtheta)= S^{a_i,b_i} \; \E_{\vtheta}\left[ \sum_{j=1}^n e^{-\int_0^{T_j} r_s ds}  \int_{T_{j-1}}^{T_j} \left(b-a-\Tra[t]{a}{b}\right)dt \right], 
\quad V^{a_i,b_i}_\text{prot}(\vtheta) = \E_{\vtheta}\left[ \int_0^T e^{-\int_0^t r_s ds}d\Tra[t]{a_i}{b_i} \right],
\end{align*}
with $S^{a_i,b_i}$ the tranche spread, the time-$t$ risk-free rate $r_t$, and the risk-free bond price $B(t)$ with maturity $t$ and notional equal to one.
In practice, we fix a time grid $0=t_0<t_1<\cdots<t_m=T$ with constant and small time step $t_i-t_{i-1}=\Delta t$ and approximate the leg values as follows,
\begin{align*}
V^{a_i,b_i}_\text{prem}(\vtheta)&\approx S^{a,b} \; \sum_{j=1}^n B(T_j)\left( (T_j-T_{j-1})(b-a)-\sum_{T_{j-1}\le t_k\le T_j} \Delta t \,E_{\vtheta}\left[ \Tra[t_k]{a}{b}\right] \right),\\
V^{a_i,b_i}_\text{prot}(\vtheta) &\approx \sum_{j=1}^m B\left(\frac{t_j+t_{j-1}}{2}\right)\left(\E_{\vtheta}\left[\Tra[t_j]{a}{b} -\Tra[t_{j-1}]{a}{b} \right]\right),
\end{align*}
where the discretization assumes that independent short-rate and the default time, as in~\cite{Mortensen2006}.

Assuming $r_t=0$ and a homogeneous portfolio with no recovery (i.e., $\ell_j = 1$), we let the default probability be $p_{j,t} = 1- e^{-\lambda t},\, j \in \left\{1, \dots, 125 \right\}$ where $\lambda$ is the credit index swap spread.
When using Beta-binomial recoveries, we use $p_{j,t} = 1- e^{-\lambda t/(1-R)}$, where $R$ is the expected recovery. 
Note that we could use individual spreads instead of the index spread. 
Because we have no access to such data, and, as will be shown in the next section, good calibration to tranches data does not warrant it, such an approach is left as direction of future research.

Finally, the model is calibrated by minimizing the squared pricing error, that is
\begin{equation} \label{eq:calibration}
\widehat{\vtheta} = \underset{\displaystyle \vtheta \subseteq \Theta}{\mbox{argmin}} \sum^4_{i=1} \left(P^{a_i,b_i}-P^{a_i,b_i}(\vtheta)\right)^2.
\end{equation}
We solve \eqref{eq:calibration} in two steps. First, we explore the parameter space to find a good starting value via a differential evolution algorithm. Second, we use the Nelder-Mead algorithm to refine the solution, enforcing the bounds by means of a parameter transformation.

\subsection{Results}
\begin{figure}
\begin{center}
\includegraphics{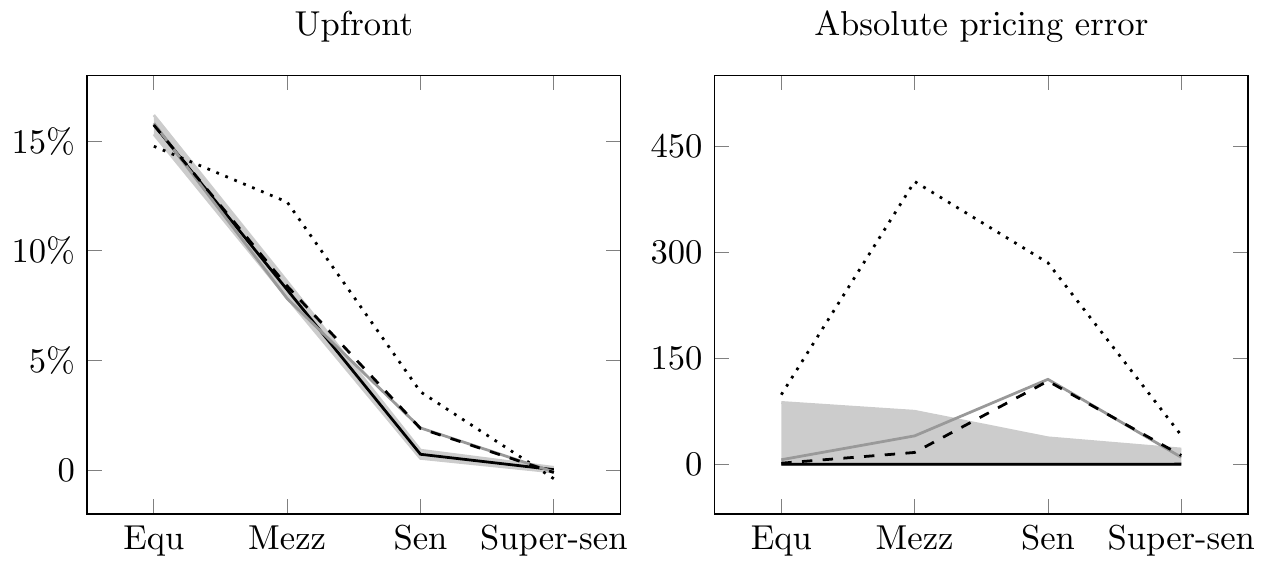}
\end{center}
\caption{CDX.NA.IG.21 upfront fits (left) in percents and absolute errors (right) in basis points on January 6th, 2014.\label{fig:CDXfits}}
\footnotesize
Fits for the one-factor Gaussian (dotted), the one-factor $t$ copula (dashed), the two-factors Gaussian-Clayton copula (grey), and the one-factor mixture with two Gaussians (black). The shaded area is the bid-ask spread.
\end{figure}

\Cref{fig:CDXfits} displays the fitted upfronts for various copulas models on January 6th, 2014. While the one-factor Gaussian (dotted line) is completely off, both the one-factor $t$ copula (dashed line) and the two-factors Gaussian-Clayton copula (grey line) perform better but miss the senior tranche. The only model achieving a perfect fit (i.e., the black line) is the following one-factor two-Gaussians mixture
\[
C_{U_j,V}(u_j,v) = w C^{\rho_1}_{U_j,V}(u_j,v) + (1-w)C^{\rho_2}_{U_j,V}(u_j,v),\quad j = 1, \dots, 125  
\]
with $w\in[0,1]$ and $C^{\rho_i}$ is a Gaussian copula with parameter $\rho_i$ (i.e., $\vtheta = (w, \rho_1, \rho_2)$ and $\Theta = [0,1] \times [-1,1] \times [-1,1]$).

\begin{figure}
\begin{center}
\includegraphics{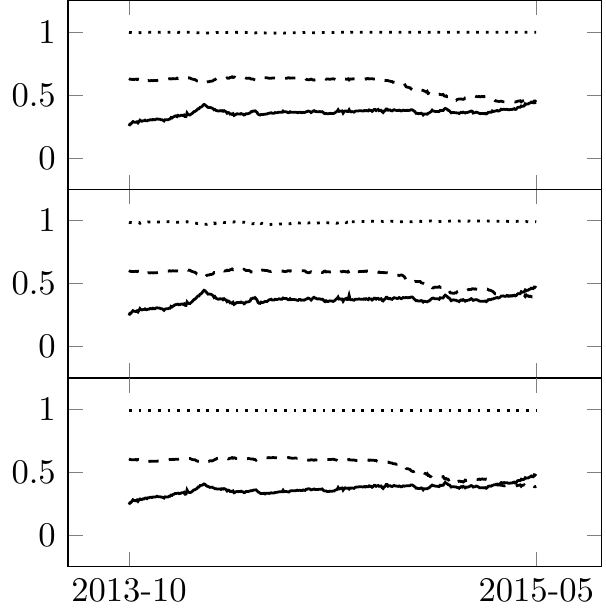}
\end{center}
\caption{CDX.NA.IG.21 time series of Gaussian mixture parameters.\label{fig:CDXparams}}
\footnotesize
Displays the calibrated $w$ (plain), $\rho_1$ (dashed), and $\rho_2$ (dotted). Three parameter sets: $\vtheta_1 = (w, \rho_1, \rho_2)$ with Beta-binomial recovery (top), $\vtheta_2 = (w, \rho_1, \rho_2)$ with zero recovery (middle) and $\vtheta_2 = (w, \rho_1, 0.99)$ with zero recovery (bottom).
\end{figure}

Repeating \eqref{eq:calibration} for the mixture models each day of the sample, we obtain time-series of calibrated parameters that we display as the plain lines in \Cref{fig:CDXparams}. There are two interesting observations that can be made. First, the parameters do not vary much over time, which indicates that the model is not over-parametrized and can be reliably estimated. Second, whether we assume a Beta-binomial or zero recovery, the parameters are very similar. Third, the parameter $\rho_2$ is close to 1 whether the recovery is zero or Beta-binomial, which means the second component of the mixture describes a comonotonic relationship between the factor and the uniform random variables for each entity. 
By fixing $\rho_2 = 0.99$ and calibrating $w$ and $\rho_1$, we obtained similar fits in terms of parameter values, with corresponding parameters time series also reported in \Cref{fig:CDXparams}.

\begin{figure}
\begin{center}
\includegraphics{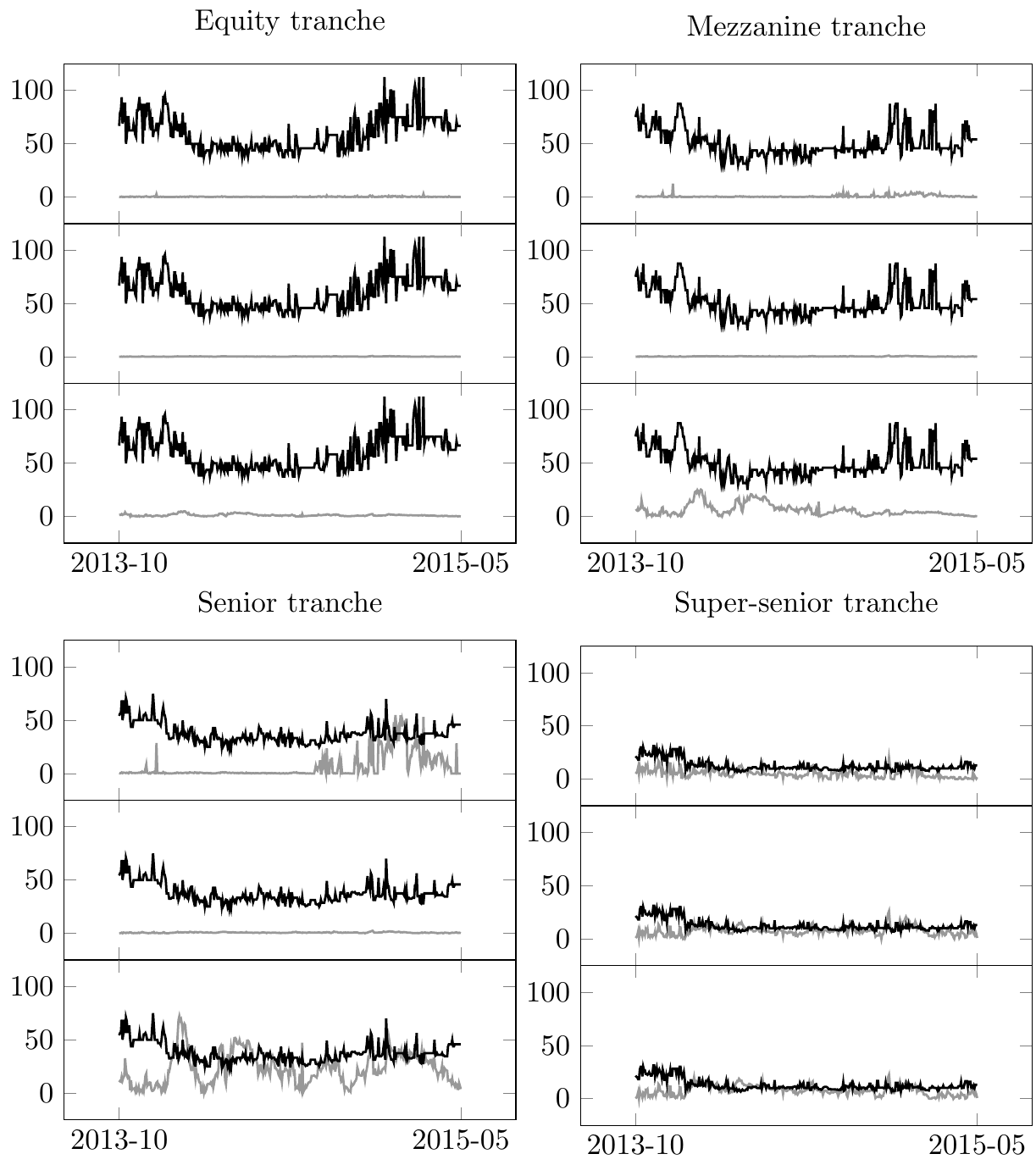}
\end{center}
\caption{Diagnostic of models calibrated on CDX.NA.IG.21 tranches.\label{fig:CDXdiagnostic}}
\footnotesize
Time-series are displayed  in basis points with the bid-ask spread (black) and the pricing errors (grey) for the model with either three parameters $(w, \rho_1, \rho_2)$ and Beta-Binomial (top) or zero (middle) recovery, or two parameters $(w, \rho_1, 0.99)$ and zero-recovery  (bottom).
\end{figure}

\Cref{fig:CDXdiagnostic} displays a model diagnostic for each of the four tranches. For each day in the sample period the pricing errors, namely $P^{a_i,b_i}-P^{a_i,b_i}(\vtheta_i)$ with $i=1,2$, are displayed as well as the bid-ask spread, $P^{a_i,b_i}_{ask}-P^{a_i,b_i}_{bid}$.
As the pricing errors are much lower than the bid-ask spread, the equity and mezzanine tranches are perfectly calibrated by both models. For the senior tranche with $\rho_2 = 1$, the end of the sample for the senior tranche and Beta-binomial recovery, and the super-senior tranche however, the pricing errors and the bid-ask spread have the same order of magnitude. To alleviate this issue, we could switch the target of the minimization in the right-hand side of \eqref{eq:calibration} from percentage of the tranche width to dollar amount. In other words, by weighting each term of the sum by $(b_i-a_i)^2$, we would increase the relative importance of the super-senior tranche in the objective function. Nonetheless, the pricing error (and the bid-ask spread) are between 10 and 30 times smaller than the upfront itself. 

To summarize, we achieve an almost perfect calibration to all tranches with only two parameters that remain stable over time. Furthermore, the assumptions on the recovery distribution have only a small impact on the overall fit quality. Using this model, one could then price tranches with non-standard attachment and detachment points, or study the impact of some distressed companies using \Cref{prop:fcmdens}.

\section{Conclusion}\label{sec:ccl}

We described a flexible and tractable class of copula-based models for dependent default times and losses.
We showed that common standard models are nested as special cases, and that many other models can easily be constructed. 
With losses taking values on a finite grid, we show how to efficiently compute the exact loss distribution of portfolios using discrete Fourier transform techniques. This allows us to study without simulations the loss distribution of complex portfolios such as portfolio of tranches, also known as CDO-squared.
We numerically study how various model features affect the portfolio loss distribution.
We calibrate multiple models to credit index tranche prices and show that a particular specification achieve almost perfect calibration to all tranches using only two parameters which appear stable over time.
Our framework is therefore a reliable solution for risk-management and pricing applications.
Potential future research directions include the estimation of bottom-up models using firm-level data, and the exploration of models with stochastic default intensities.

\newpage
\appendix

\section{Proofs}\label{sec:proofs}

This Appendix contains the proofs of all theorems and propositions in the main text.

\subsection*{Proof of \Cref{lem:jointDefPro}}

The joint probability of default rewrites
\[
\Pro{ \tau_1 \le t_1 ,\dots, \, \tau_N \le t_N } = \Pro{ U_1 \le p_{1,t_1},\dots, \, U_N \le p_{N,t_N} }= C_U\left(p_{1,t_1} ,\dots, \, p_{N,t_N} \right)
\]
where the second line follows by definition of $C_U$.

\subsection*{Proof of \Cref{prop:1fcm}}

Observe that for all $j=1,\dots,\,N$ the random vector $(U_j,V)$ takes values on $[0,1]^2$ and has uniform marginal densities, this implies that
\[
\Pro{ U_j \le u_j, V \le v} = C_{U_j,V}(u_j,v)
\]
for some bivariate copulas $C_{U_j,V}$ and any $(u_j,v)\in[0,1]^2$.
Therefore we have
\[
\Pro{ U_j \le u_j \mid V = v} = C_{U_j\mid V}(u_j\mid v)
\]
and by plugging this into Equation~\eqref{eq:1fcm} then integrating with respect to the density $f_V(v)=v$ of $V$ we obtain
\[
C_U(u_1,\dots,\,u_N) = \int_0^1 \prod_{j=1}^N \Pro{ U_j \le u_j, V \le v} f_V(v)dv = \int_0^1 \prod_{j=1}^N C_{U_j,V}(u_j,v) dv.
\]
The desired expression then follows from \Cref{lem:jointDefPro}.

\subsection*{Proof of \Cref{prop:fcmdens}}

We denote $U_{\rm I}$ the vector which contains the coordinates of $U$ which are in $\Ical\setminus\Dcal$, and $U_{\rm J}$ contains the ones which are in $\Dcal$.
For readability we assume that the coordinates are ordered according to $U=(U_{\rm I},U_{\rm J})$.
Similarly we group the marginal default probabilities into two vectors $p_{\rm I}$ and $p_{\rm J}$.
The size of $U_{\rm I}$ and $U_{\rm J}$ are respectively given by $N_{\rm I}$ and $N_{\rm J}=N-N_{\rm I}$.
We directly write the proof for the multivariate case with $V\in[0,1]^d$.
The joint default probability conditional on the default of the $k\in\Dcal$ entities then rewrites
\begin{equation*}
\begin{split}
\Pro{ \tau_1 \le t_1 ,\dots, \, \tau_N \le t_N \mid \tau_k=t_k:k\in\Dcal} 
 = \Pro{ U_1 \le p_{1,t_1}, \dots, \, U_N \le p_{N,t_N} \mid U_k = p_{k,t_k}:k\in\Dcal} \\
 = \frac{\int_{[0,p_{\rm I}]} c_{U_{\rm I},U_{\rm J}}(u_{\rm I},p_{\rm J})du_{\rm I} }{\int_{[0,1]^{N_{\rm I}}} c_{U_{\rm I},U_{\rm J}}(u_{\rm I},p_{\rm J})du_{\rm I}}= \frac{\int_{[0,1]^d} C_{U_{\rm I}\mid V}(p_{\rm I}\mid v) c_{U_{\rm J}, V}(p_{\rm J} , v) dC_V(v) }{\int_{[0,1]^d} c_{U_{\rm J}, V}(p_{\rm J}, v) dC_V(v)}.
\end{split}
\end{equation*}
The second equality comes from the definition of the conditional probability for measures.
The third equality follows from the definition of the factor copula and Fubini's theorem,
\[
\int_{[0,x]} \; c_{U_{\rm I},U_{\rm J}}(u_{\rm I},p_{\rm J}) \; du_{\rm I} = \int_{[0,1]^d} \; \prod_{U_j\in U_{\rm I}} \; \int_0^{x_j} c_{U_j, V}(u, v) \; du \; \prod_{U_j\in U_{\rm J}} c_{U_j, V}(u_j, v) \; dC_V(v),
\]
along with $\int_0^{x_j} c_{U_j, V}(u, v) du = C_{U_j\mid V}(x_j\mid v)$ and $C_{U_j\mid V}(1\mid v)=1$.

Alternatively, the same result can be proved by first showing that
\begin{align*}
 \Pro{ \tau_1 \le t_1 ,\dots, \, \tau_N \le t_N \mid \{\tau_k=t_k:k\in\Dcal\}\cup\{V=v\}} &= \frac{\int_{[0,p_{\rm I}]} c_{U_{\rm I}, V}(x, v)dx \;  c_{U_{\rm J}, V}(p_{\rm J}, v) \; c_V(v)}{\int_{[0,1]^{N_{\rm I}}} c_{U_{\rm I}, V}(x, v)dx \; c_{U_{\rm J}, V}(p_{\rm J}, v) \; c_V(v)}\\
&= \prod_{j\in\Ical\setminus\Dcal}C_{U_j\mid V}(p_{j,t_j}\mid v)
\end{align*}
and then integrating with respect to the following conditional density
\[
\Pa[V \le v \mid \{U_k = p_{k,t_k}:k\in\Dcal\}] = \frac{\int_{[0,v]} \prod_{j\in\Dcal} c_{U_j, V}(p_{j,t_j}, x) dC_V(x)}{\int_{[0,1]^d} \prod_{j\in\Dcal} c_{U_j, V}(p_{j,t_j}, x) dC_V(x)}.
\]

\subsection*{Proof of \Cref{prop:fcm}}

The $V$-conditional joint default probability as a similar expression as in Equation~\eqref{eq:VcondJDP}.
The unconditional joint default probability follows by integrating with respect to the joint density $c_V(v)$ of $V$ which gives the expression for $C_U$ as $dC_V(v)=c_V(v)dv$.
Observe now that the joint distribution of the random vector $(U_j,V)$ is by construction given by a $(1+d)$-dimensional copula $C_{U_j,V}$ for all $j\in\Ical$.
By definition we must have
\begin{align*}
C_{U_j,V} (u_j,v) &= \Pro{U_j\le u_j, \,  V \le v}= \int_0^{v_1}\dots\int_0^{v_d} \Pro{U_j\le u_j \mid V=y} d\Pro{V\le y}\\
&= \int_0^{v_1}\dots\int_0^{v_d} C_{U_j\mid V} (u_j\mid y) dC_V(y)
\end{align*}
for all $(u_j,v)\in[0,1]^{1+d}$ which gives Equation~\eqref{eq:fcm}.

\subsection*{Proof of \Cref{cor:fcmPCC}}
The density of $V$ is given by $C_V(v) = \prod_{j=1}^d v_j$, and following \cite{joe1996} the conditional copulas are given by
\begin{align*}
 C_{U_j \mid V}(u_j\mid v) = \frac{\partial C_{U_j,V_k \mid V_{-k}}\left( C_{U_j\mid V_{-k}}(u_j\mid v_{-k}), v_k \mid v_{-k} \right)}{\partial v_k}
\end{align*} 
for any $k =  1, \dots,\, d$, and where $V_{-k} = (V_{1}, \dots, V_{k-1},V_{k+1},\dots, V_d)$ denotes the random vector $V$ without its $k$-th coordinate.
By iterating the previous equation, the conditional copula $C_{U_j \mid V}(u_j\mid v) $ can be rewritten as a recursive composition of bivariate linking copulas
\begin{equation*}
C_{U}(u_1,\,\dots,\,u_n) = \int_{[0,1]^d} \prod_{j=1}^N C_{U_j\mid V_1}(\cdot|v_1) \circ \cdots \circ C_{U_j\mid V_d}(u_j|v_d) dv
\end{equation*}
where $C_{U_j, V_k}$ denotes a bivariate copula for $j=1,\,\dots,\,N$ and $k=1,\,\dots,\,d$.

\subsection*{Proof of \Cref{thm:fm2fcm}}

Observe that the random vector $U=(F_{Y_1}(Y_1),\,\dots,\,F_{Y_N}(Y_N))$ and $V=(F_{X_1}(X_1),\,\dots,\,F_{X_d}(X_d))$ have uniform margins by construction suggesting that their distributions are given by copulas.
The following theorem proves the existence of $C_V$.
\begin{theorem}[Sklar's Theorem 1959]
$F_V$ is a joint distribution with margins $F_{X_i}$ for $i \in \left\lbrace 1, \cdots, d \right\rbrace $ if and only if there exists a copula $C_V$, that is a distribution which is supported in the unit hypercube and has uniform margins, such that
\begin{equation*}
F_X\left(x_1,\,\dots,\,x_N \right) = C_V\left(F_{X_1}(x_1),\,\dots,\,F_{X_N}(x_N) \right)
\end{equation*}
for all $x\in\R^N$. 
Moreover, if the margins are continuous, then $C_V$ is unique.
\end{theorem}
For all $v\in [0,1]^d$ the theorem implies that
\begin{align*}
C_V\left( v_1 , \, \dots, \, v_d \right) &=
F_X\left( F^{-1}_{X_1}(v_{1}), \, \dots, \,F^{-1}_{X_d}(v_{d})  \right) = \Pa\left[ X_1 \leq F^{-1}_{X_1}(v_{1}), \, \dots, \, X_d \leq F^{-1}_{X_d}(v_{d}) \right]\\
&= \Pa\left[ F_{X_1}(X_1) \leq  v_1, \, \dots, \, F_{X_d}(X_d) \leq v_d \right]  = \Pa\left[ V_1 \leq  v_1, \, \dots, \, V_d \leq v_d \right] .
\end{align*} 
The copula $C_V$ is thus the joint distribution of probability integral transforms.
The $X$-conditional independence of $Y$ implies that
\begin{equation*}
\Pa\left[ Y_1 \leq y_{1t_1}, \, \dots, \, Y_N \leq y_{Nt_N} \mid X = x \right] = \prod^N_{j=1} F_{Y_j \mid X}(y_{jt_j} \mid x),
\end{equation*}
where $F_{Y_j \mid X}$ denotes the distribution of $Y_j$ conditional on $X$ such that 
\begin{equation*}
\Pro{\tau_j \leq t_j \mid X = x} = \Pro{U_j \leq p_{j,t_j} \mid V = v},
\end{equation*}
where $v=\tilde{F}_X(x):=\left( F_{X_1}(x_{1}), \, \dots, \,F_{X_d}(x_{d})  \right)$.
A copula representation of the above probability can finally be obtained by applying the conditional equivalent of Sklar's theorem:
\begin{theorem}[Patton's Theorem 2002]
$F_{Y \mid X}$ is a joint conditional distribution with conditional margins $F_{Y_i \mid X}$ for $i \in \left\lbrace 1, \cdots, N \right\rbrace $ if and only if there exists a conditional copula $C_{U \mid V}$, that is a conditional distribution which is supported in the unit hypercube and has uniform conditional margins, such that
\begin{align*}
F_{Y \mid X}\left( y_1, \dots, y_N \mid x  \right) &= C_{U \mid V}(F_{Y_1 \mid X}(y_1 \mid x), \dots, F_{Y_N \mid X}(y_N \mid x) \mid F_X(x))
\end{align*}
for all $y\in\R^N$ and $x \in \R$. Moreover, if the conditional margins are continuous, then $C_{U \mid V}$ is unique.
\end{theorem}
For all $u\in [0,1]^N$ and $v \in [0,1]^d$ the theorem implies
\begin{align*}
C_{U\mid V}\left( u_1 , \, \dots, \, u_N \mid v \right) &=
F_{Y\mid X}\left( F^{-1}_{Y_1 \mid X }(u_{1}), \, \dots, \, F^{-1}_{Y_N\mid X}(u_{N}) \mid \tilde{F}^{-1}_X(v) \right)\\
&= \Pa\left[ Y_1 \leq F^{-1}_{Y_1\mid X}(u_{1}), \, \dots, \, Y_N \leq F^{-1}_{Y_N\mid X}(u_{N}) \mid X = \tilde{F}^{-1}_X(v) \right] \\
&= \Pa\left[ F_{Y_1\mid X}(Y_1\mid X) \leq  u_N, \, \dots, \, F_{Y_N\mid X}(Y_N\mid X) \leq u_N \mid \tilde{F}_X(X) = v \right]\\
&= \Pa\left[ U_1 \leq  u_1, \, \dots, \, U_N \leq u_N \mid V = v \right].
\end{align*} 
In other words, the copula $C_{U \mid V}$ is also the joint conditional distribution of the conditional probability integral transforms. 
As such, the joint conditional distribution of default times is given by
\begin{align*}
\Pa\left[ \tau_1 \leq t_1, \, \dots, \, \tau_N \leq t_N \mid X = F^{-1}_X(v) \right] &= \Pa\left[ U_1 \leq   p_{1,t_1}, \, \dots, \, U_N \leq p_{N,t_N} \mid V = v\right]\\
&= C_{U \mid V}\left( p_{1,t_1} , \, \dots, \, p_{N,t_N} \mid v \right),
\end{align*}
which completes the proof.

\subsection*{Proof of \Cref{prop:CICFformula}}

The default times and the loss amounts being independent conditional on $V$ we have
\[
\cExp{ \e^{\im u L_t \delta^{-1}}}{V=v} = \cExp{ \e^{\im u \sum_{j=1}^N \Ind{\tau_j\leq t} \ell_j \delta^{-1}}}{V=v} = \prod_{j=1}^N \cExp{ \e^{\im u \Ind{\tau_j\leq t} \ell_j \delta^{-1} }}{V=v}
\]
Furthermore, by independence of the random variables $\Ind{\tau_j\leq t}$ and $ \ell_j$ conditional on $V$ we have
\[
\cExp{ \e^{\im u \Ind{\tau_j\leq t} \ell_j \delta^{-1} }}{V=v} =  1 - \cPro{\tau_j\leq t}{V=v}  + \cPro{\tau_j\leq t}{V=v} \phi_{\ell_j}(u,v)
\]
where $\phi_{\ell_j}(u,v):=\cExp{\e^{\im u \ell_j \delta^{-1}}}{V=v}$ denotes the $V$-conditional characteristic function of $\ell_j\delta^{-1}$.
We finally apply the tower property
\begin{align*}
\phi_{L_t}(u) &= \Exp{ \cExp{ \e^{\im u L_t \delta^{-1}}}{V=v} } = \int_{[0,1]^d} \cExp{ \e^{\im u L_t \delta^{-1}}}{V=v} dC_V(v)\\
&=\int_{[0,1]^d} \left(1-p_{j,t}(v)+p_{j,t}(v)\phi_{\ell_j}(u,v) \right) dC_V(v)
\end{align*}
where $C_V$ is the density of $X$, and $p_{j,t}(v)=C_{U_j\mid V}(p_{j,t}\mid v)$.

\subsection*{Proof of \Cref{lem:cf2pmf}}

The proof is an application of discrete Fourier transform inversion.
Observe that the random variable $L_t\delta^{-1}$ has state space $\{0,\,1,\,\dots,\,M\}$.
Its discrete Fourier transform is given by
\begin{align*}
F_m &= \sum_{k=0}^M \Pro{L_t\delta^{-1}=k} \e^{-\im\frac{2\pi m k}{M+1}} =\phi_{L_t}\left( \frac{-2\pi m}{(M+1)}\right)
\end{align*}
where $\phi_{L_t}$ as in \Cref{prop:CICFformula} is the characteristic function of $L_t \delta^{-1}$.
The probability mass function can be recovered as follows
\begin{align*}
\Pro{L_t=k\delta} &= \frac{1}{M+1} \sum_{m=0}^M F_m \e^{\im\frac{2\pi m k}{M+1}}.
\end{align*}
Equation~\eqref{eq:cf2pmf} follows by observing that the signs can equivalently be switched between the complex weights.

\subsection*{Proof of \Cref{prop:tranche}}

The proof of this proposition is immediate from the factor copula construction with discretely supported on a finite grid.

\subsection*{Proof of \Cref{cor:CDOsq}}

This follows directly from \Cref{prop:CICFformula} and  \Cref{prop:tranche}.

\subsection*{Proof of \Cref{prop:NtLtpmf}}
By construction we have
\begin{align*}
F_{x,y} : &= \phi_{N_t,L_t}(\mu x,\,\nu y)  = \E\left[\prod_{j=1}^N \exp\left\{ \im \, \Ind{\tau_j \le t} (\mu x + \nu y \ell_j \delta^{-1}) \right\} \right] \\
& = \E\left[\exp\left\{\sum_{j=1}^N  \im \, \Ind{\tau_j \le t} (\mu x + \nu y \ell_j \delta^{-1}) \right\}  \right]  = \E\left[ e^{ \im\mu x N_t + \im\nu y L_t \delta^{-1} }  \right]
\end{align*}
Using this last expectation and the explicit expressions for $\mu$ and $\nu$ we obtain
\begin{align*}
F_{x,y} &  = \sum_{j=0}^N \sum_{k=0}^M \Pro{N_t=j,L_t=\delta k} e^{ \im \frac{2\pi j}{N+1}x} e^{ \im \frac{2\pi k}{M+1}y}.
\end{align*}
This last expression is the two dimensional discrete Fourier transform of the density of the variable $(N_t,L_t\delta^{-1})$.
The density can then immediately be retrieved by applying the inverse two-dimensional discrete Fourier transform inversion as follows
\[
\Pro{N_t=j,L_t=\delta k} =  \sum_{x=0}^N \sum_{y=0}^M F_{x,y} \, e^{ - \im \frac{2\pi x}{N+1}j} e^{ -\im \frac{2\pi y}{M+1}k}.
\]

\section{Standard Copula Models}\label{sec:sfm}

We derive in this section the factor copula representation of the most popular models that have been proposed in the literature on multi-name credit risk.

\paragraph*{Gaussian copula models.}
Let us denote the Gaussian copula and $h$-function by
\begin{align*}
C_{U,V}^{G}(u,v;\rho)&=\Phi_2\left(\Phi^{-1}(u),\Phi^{-1}(v);\rho \right) 
\end{align*}
and
\begin{align*}
C_{U ; V}^{G}(u \mid v;\rho) &= \Phi\left(\frac{\Phi^{-1}(u)-\rho \Phi^{-1}(v)}{1-\rho^2} \right),
\end{align*}
where $\Phi(\cdot)$ is the standard normal distribution and $\Phi_2(\cdot,\cdot;\rho)$ is the bivariate normal distribution with correlation $\rho$. For instance, when $d=1$ and all bivariate copulas are Gaussian,  then a representation for the joint distribution of default times is the copula of a $1$-factor model
\begin{align*}
Y_j = \beta_{j} X + \sqrt{1-\beta_{j}^2} Z_j,
\end{align*}
where $X,Z_1, \dots, Z_N$ are i.i.d. $N(0,1)$ random variables. In this case, the correlation parameter for the bivariate copula linking the default of obligor $j$ to the systematic factor is $\beta_{j}$.  By considering a unique correlation parameter $\beta_j = \rho$ for $j \in \left\{1, \dots,N \right\}$, \cite{li2000default} is a special case of our formulation. Furthermore, when $d>1$, then a representation for the joint distribution of default times is the copula of a $d$-factor model
\begin{align*}
Y_j = \sum^p_{i=1} \beta_{j,i} X_i + Z_j,
\end{align*} 
where $X_1,\dots,X_p,Z_1, \dots, Z_N$ are i.i.d.\ $N(0,1)$ random variables. In this case, the parameters for the second to $d$ factors are partial correlations, namely
 \[
\rho_{U_j,V_k \mid X_1, \dots, V_{k-1}} = \frac{Cov(Y_j,X_k \mid X_1,\dots,X_{k-1})}{\sqrt{Var(Y_j \mid X_1,\dots,X_{k-1})}\sqrt{Var(X_k \mid X_1,\dots,X_{k-1})}} = \frac{\beta_{j,k}}{\sqrt{1-\beta_{j,1}^2 - \dots - \beta^2_{j,k-1}}}.
\]

\paragraph*{Stochastic correlation models.}

It is straightforward to build more complex factor models, stochastic correlations models are obtained by writing
\begin{align*}
Y_j = \left(B_j\alpha_j + (1-B_j)\beta_j \right) X + \sqrt{1-\left(B_j\alpha_j + (1-B_j)\beta_j \right)^2} Z_j,
\end{align*}
where $B_j$ are i.i.d. Bernoulli$(b_j)$ and $X,Z_1, \dots, Z_N$ as before. For this model, the bivariate copulas are convex sum of Gaussian copulas, that is
\begin{align*}
C_{U_j,V}^{SC}(u_j,v;\alpha_j,\beta_j,b_j) &= b_j C_{U, V}^{G}(u , v;\alpha_j)+(1-b_j) C_{U, V}^{G}(u , v;\beta_j).
\end{align*}
Deriving the $h$-function is then straightforward.

\paragraph*{The $t$-Student model.}
Usually, $t$-student models are specified by considering, 
\begin{align*}
Y_j = \sqrt{W}\left(\beta_{j} X + \sqrt{1-\beta_{j}^2} Z_j \right)
\end{align*}
where $W$ is an i.i.d. random variable such than $\nu/W$ is $\chi^2(\nu)$ and $X,Z_1, \dots, Z_N$ as before. Then the default times are independent conditional on $(W,X)$ and their conditional probability distribution is easily derived (see e.g. \cite{Burtschell2009}). Using our formulation, we obtain an equivalent $t$-student model by considering the copula and $h$-function directly, that is
\begin{align*}
C_{U,V}^{t}(u,v;\rho,\nu) &= t_2(t^{-1}_{\nu}(u),t^{-1}_{\nu}(v);\rho,\nu)
\end{align*}
and 
\begin{align*}
C_{U ; V}^{t}(u,v;\rho,\nu) = t_{\nu+1} \left( f(u,v) \right), \mbox{ with } f(u,v)=\frac{t^{-1}_{\nu}(u) - \rho t^{-1}_{\nu}(v)}{\sqrt{\frac{(1-\rho^2) \left(\nu+ \left( t^{-1}_{\nu}(v)\right)^2\right) }{\nu + 1}}},
\end{align*}
where $t_{\nu}(\cdot)$ is the $t$-student distribution with $\nu$ degrees of freedom and $t_2(\cdot,\cdot;\rho,\nu)$ is the bivariate $t$-student distribution with correlation $\rho$ and degrees of freedom $\nu$. Compared to the other formulation, our alternative only require a one-dimensional integration. Furthermore, using different degrees of freedom for each bivariate copulas offers additional modeling flexibility without additional cost. 

\paragraph*{Archimedean models.}
One-parameter archimedean copulas are built by considering a continuous, strictly decreasing and convex generator $\psi: [0,1] \times \Theta \rightarrow [0,\infty) $ such that $\psi(1;\theta)=0$ for all $\theta \in \Theta$, where $\Theta$ represents the parameter space. Using this generator, a bivariate copula is obtained by writing
\begin{align*}
C^{\psi}_{U,V}(u,v;\theta) = \psi^{-1} \left(\psi(u;\theta)+\psi(v;\theta);\theta \right).
\end{align*}
For such a copula, the $h$-function $C^{\psi}_{U ; V}$ is usually straightforward to derive, and we summarize the most popular in \Cref{table:archimedean}.

\begin{table}
\begin{center}
\begin{tabular}{c|c|c|c}
 \hline
 & Generator $\psi$ & Inverse generator $\psi^{-1}$ & Parameter space $\Theta$  \\ \hline 
Clayton &  $\frac{u^{-\theta}-1}{\theta}$& $(1+\theta u)^{-1/\theta}$ & $(0,\infty)$ \\ 
Gumbel & $\left(- \log(u) \right)^{\theta}$ & $\exp \left( - u^{1/\theta} \right)$ &  $[1,\infty)$ \\ 
Frank & $- \log \left( \frac{ \exp \left( - \theta u \right) - 1}{\exp \left( - \theta \right) - 1} \right)$ & $- \frac{1}{\theta} \log \left( 1 + \exp \left( -t \right) \left( \exp \left( - \theta \right) - 1 \right)  \right)$ & $(-\infty,\infty) \setminus \left\{ 0 \right\}$  \\ 
Joe & $- \log \left( 1 - (1-u)^{\theta} \right)$ &  $1-\left( 1 - \exp (-u) \right)^{1/\theta}$ & $[1,\infty)$ \\ 
Independence & $- \log (u)$ & $\exp (-u) $ &  $\varnothing$ \\
 \hline
\end{tabular}

\vspace{0.5cm}

\begin{tabular}{c|c}
 \hline
& Copula $C^{\psi}_{U,V}$ \\ \hline
Clayton &  $\left(u^{-\theta}+v^{-\theta}-1 \right)^{-1/\theta}$ \\ 
Gumbel &  $e^{- \left( \left(- \log(u) \right)^{\theta} + \left(- \log(v) \right)^{\theta} \right)^{1/\theta}}$\\
Frank & $- \frac{1}{\theta} \log \left( \frac{1-e^{-\theta} - \left(1-e^{-u \theta} \right) \left(1-e^{-v \theta} \right)}{1-e^{-\theta}}   \right)$  \\
Joe & $1 - \left( (1-u)^{\theta} + (1-v)^{\theta} - (1-u)^{\theta} (1-v)^{\theta} \right)^{1/\theta}$  \\
Independence & $uv$  \\
 \hline
\end{tabular}

\vspace{0.5cm}

\begin{tabular}{c|c}
 \hline
 & $h$-function $C^{\psi}_{U \mid V}$\\ \hline
Clayton  &  $C^{\psi}_{U,V}(u,v;\theta) v^{-1-\theta}$\\ 
Gumbel  &  $ C^{\psi}_{U,V}(u,v;\theta) \frac{\left( \left(- \log(u) \right)^{\theta} + \left(- \log(v) \right)^{\theta} \right)^{1/\theta - 1} \left(- \log(v) \right)^{\theta}}{v \log(v)} $\\
Frank  & $\frac{\displaystyle e^{\theta} \left( e^{\theta u} -1 \right)}{\displaystyle  e^{\theta u + \theta v} - e^{\theta u + \theta }  - e^{\theta v + \theta } + e^{\theta v} }$ \\
Joe  & $\left(C^{\psi}_{U,V}(u,v;\theta)\right)^{1-\theta} (1-v)^{\theta - 1} (1-(1-u)^{\theta})$ \\
Independence & $u$ \\
 \hline
\end{tabular}
\end{center}
\caption{Archimedean copulas}
\scriptsize
Describes the generator $\psi$, the inverse generator $\psi^{-1}$, the parameter space $\Theta$, the copula $C^{\psi}_{U,V}$, and the $h$-function $C^{\psi}_{U \mid V}$.
\label{table:archimedean}
\end{table}

\newpage
\bibliographystyle{alpha}
\bibliography{../copula_cdo}

\end{document}